\documentclass[final,5p,times,twocolumn]{elsarticle}

\usepackage{graphicx}
\usepackage[amssymb,mediumqspace,thinspace]{SIunits}
\usepackage{xcolor}
\usepackage[normalem]{ulem}

\usepackage{lineno}

\biboptions{sort&compress}
\modulolinenumbers[5]
\journal{International Journal of Plasticity}
\begin{document}

\begin{frontmatter}

\title{Dislocation interactions and crack nucleation in a fatigued near-alpha titanium alloy}



\author{S. Joseph}
\author{T.C. Lindley}
\author{D. Dye\corref{mycorrespondingauthor}}
\cortext[mycorrespondingauthor]{Corresponding author}
\ead{david.dye@imperial.ac.uk}

\address{Department of Materials, Royal School of Mines, Imperial College, Prince Consort Road, South Kensington, London SW7 2BP, UK}

\begin{abstract}
Dislocation interactions at the crack nucleation site were investigated in near-$\alpha$ titanium alloy Ti-6242Si subjected to low cycle fatigue. 
Cyclic plastic strain in the alloy resulted in dislocation pile-ups in the primary $\alpha$ grains, nucleated at the boundaries between the primary $\alpha$ and the two-phase regions. These two phase regions provided a barrier to slip transfer between primary alpha grains.
We suggest that crack nucleation occurred near the basal plane of primary $\alpha$ grains by the subsurface double-ended pile-up mechanism first conceived by Tanaka and Mura. Superjogs on the basal $\langle a \rangle$ dislocations were observed near the crack nucleation location.
The two phase regions showed direct transmission of $a_3$ dislocations between secondary alpha plates, transmitted through the $\beta$ ligaments as $a[010]$, which then decompose into $(a/2)\langle 111 \rangle$ dislocation networks in the $\beta$. The $\beta$ ligaments themselves do not appear to form an especially impenetrable barrier to slip, in agreement with the micropillar and crystal plasticity investigations of Zhang et al.
\end{abstract}


\end{frontmatter}


\section{Introduction}
Near-$\alpha$ titanium alloys such as Ti-6242Si are used in elevated temperature service in aero-engines, e.g. in the compressor, especially in the 400--$550\celsius$ range. A wide range of temperature, stress and loading frequency combinations are experienced across the flight cycle; for example during take-off stresses in the disc bore can be high before the thermal field has equilibrated, and so room temperature fatigue behaviour can be an important consideration. Titanium alloys are generally notch sensitive and so crack initiation from the microstructure is often of greater concern than in other alloy systems where lifing can mostly be considered in terms of crack growth.  

Extensive research on fatigue crack initiation mechanisms have been performed, including metallographic observations of crack initiation \cite{kim1978crack, luquiau1997cyclic, chan1981deformation, wagner1987microstructural, boyer1993microstructure, dunne2007experimental, tan2015effect}, intrusions and extrusions on the sample surface \cite{bao1989overview, basinski1992fundamental, ahmed2001electron, polak2016experimental}  and observations of dislocations structures by TEM~\cite{basinski1969early, essmann1979annihilation, laird1986low, basinski1992fundamental, beranger1993low, pedersen1995cyclic, polak2016experimental}. Based on these findings, theoretical models of fatigue crack nucleation ~\cite{mott1958theory, essmann1981model, tanaka1981dislocation, sangid2011physically, polak2016experimental} have been proposed to explain the mechanism of fatigue crack nucleation. Essmann et al~\cite{essmann1981model} proposed the first such model to explain the intrusions and extrusions formed. Tanaka and Mura~\cite{tanaka1981dislocation} then modeled the formation of an embryonic crack from the dislocation pile-ups accumulated during cyclic loading. 
	
Ti-6242Si develops a wide variety of microstructures depending upon the thermo-mechanical history, with microstructure having a significant effect on fatigue behaviour~\cite{singh2002low, xiao2002cyclic, singh2007low, li2007comparison, huang2011cyclic, wu2013effect}. It is also evident that the stress evolution in dual phase titanium alloys is mainly governed by the primary alpha $\alpha_p$ phase~\cite{luquiau1997cyclic, chan1981deformation}, because the $\alpha_p$ is usually softer than the beta phase, $\beta$. 

\textcolor{black}{However, the two-phase regions having thin secondary alpha ($\alpha_s$) plates in retained $\beta$ also play a role. The two-phase region was found to be a barrier for slip transmission between primary alpha grains during low cycle fatigue \cite{joseph2018slip}. The slip transmission within this two-phase region is mainly dependent on the orientation relation between the $\alpha_s$ and $\beta$ plates. In general, it follows the Burgers orientation relation \cite{bhattacharyya2003role}, however, other crystallographic variants are also possible \cite{tong2017using}.  Slip transmission has been reported between $\alpha$ and $\beta$ plates \cite{savage2004anisotropy,zhang2016determination} and it was more likely to be observed when the $\beta$ plates are favorably oriented for slip \cite{seal2012analysis}.} 

In lamellar microstructures fatigue cracks initiate either within slip bands in $\alpha$ lamellae or along prior beta grain boundaries~\cite{wagner1987microstructural, boyer1993microstructure}.  In bimodal microstructures, cracking can initiate either on the interface between the lamellar microstructure and the $\alpha_p$ or within the $\alpha_p$ itself; initiation also depends on the fraction and size of the $\alpha_p$. In near-$\alpha$ alloys, such as Ti6242, with more than 50\% $\alpha_p$, crack nucleation is expected to occur in the $\alpha_p$ due to extensive grain boundary dislocation pile-up~\cite{tan2015effect}. Further, a recent in-situ micro-fatigue study in Ti6246 identified that crack nucleation sites are associated with microtextured regions favourably oriented for basal or prism slip~\cite{szczepanski2013demonstration}.

Despite this extensive research, very few observations have been made of the dislocation structures associated with naturally initiated fatigue crack initiation sites, as opposed to those associated with cyclic plasticity from gauge sections, or SEM studies of the orientations and coarse microstructural features associated with cracking. However, it has recently become possible, with focused ion beam milling, to perform such site-specific studies. 

In this paper, we perform such studies with the aim of elucidating the dislocation mechanisms  \textcolor{black}{near a fatigue crack origin}, in order that more fatigue-resistant microstructures and alloys might be developed in future. The dislocation mechanisms associated with fatigue crack nucleation and the role of the two-phase regions are discussed in detail.
	
\section{Experimental Description}
The Ti-6Al-2Sn-4Zr-2Mo-0.1Si (wt.\%) alloy investigated in this work was melted from elemental stock and then processed by rolling in both the $\beta$ and $\alpha+\beta$ domains, recrystallized in the $\alpha+\beta$ domain at $950\celsius$ for $5\usk\hour$ and air-cooled. The alloy was then aged at $593\celsius$ for 8h and air cooled to promote nanometre-scale Ti$_3$Al precipitation. This processing route resulted in bimodal microstructure with 50\% volume fraction of $\alpha_p$ in the transformed $\beta$.  \textcolor{black}{The initial microstructure of the alloy can be found in our previous work  \cite{joseph2018slip}}.  Very thin secondary alpha $\alpha_s$ platelets of 50--$100\usk\nano\meter$ thickness were observed in the retained $\beta$.
	 
Low cycle fatigue (LCF) tests were carried out on cylindrical plain fatigue samples with a $2.9\usk\milli\meter$ diameter and $15\usk\milli\meter$ long gauge length using a Mayes servohydraulic machine with an Instron 8800 controller. A trapezoidal waveform with a ramp up/down time of $1\usk\second$, a $1\usk\second$ hold at maximum stress of 95\% of yield stress ($831\usk\mega\pascal$ ), $1\usk\second$ hold at minimum stress and an $R$ ratio of 0.05 was used. One sample failed after 23,914 cycles.

A Zeiss Auriga field emission gun scanning electron microscope (FEG-SEM) in secondary electron imaging mode was used for fractography. Dislocation analysis was conducted using a JEOL JEM-2100F TEM/STEM with an accelerating voltage of $200\usk\kilo\volt$. TEM samples were prepared using the focused ion beam (FIB) lift-out technique in a dual beam FEI Helios NanoLab 600 using a $30\usk\kilo\volt$ Ga ion beam. Foils were extracted near the crack origin from the fracture surface. To protect the area of interest, a gas injection system was used to deposit a Pt-containing protective layer.

Samples were made electron transparent by thinning down to a thickness of 150 nm. To obtain the orientation relation between $\alpha_s$ and $\beta$ ligaments in a two phase region of the foil, Transmission Kikuchi Diffraction (TKD) was carried out. TKD data was collected on the same Auriga SEM used for fractography, at an accelerating voltage of $30\usk\kilo\volt$ and at a working distance of $2\usk\milli\meter$. The stage was tilted to $30\degree$ to make the sample in the TKD holder normal to the electron beam. Detailed description of the orientation relations found by TKD can be found in Tong et al~\cite{tong2017using}.

\section{Results}
The fracture surface was investigated and smooth, near-planar crack initiating facets were found very slightly below the sample surface, Figure 1(a).  The spatial and crystallographic orientation of these facets obtained by quantitative tilt fractography and EBSD are reported in Joseph et al.\cite{joseph2018slip}. TEM foils were extracted from the two regions identified, at the crack nucleation site and, for comparison purposes, a region $50\usk\micro\meter$ away in the fatigue crack growth region. The foils are shown in Figure 1(b-c) and the $\alpha_p$ and two phase regions ($\alpha_s+\beta$) regions marked. Each grain in the foils was tilted to at least three different beam directions $B$ and three different $g$ vectors under each beam condition to analyze the dislocations. The dislocations observed in the foils are discussed in the following sections.

\begin{figure}[t]
\begin{center}\includegraphics[width=90mm]{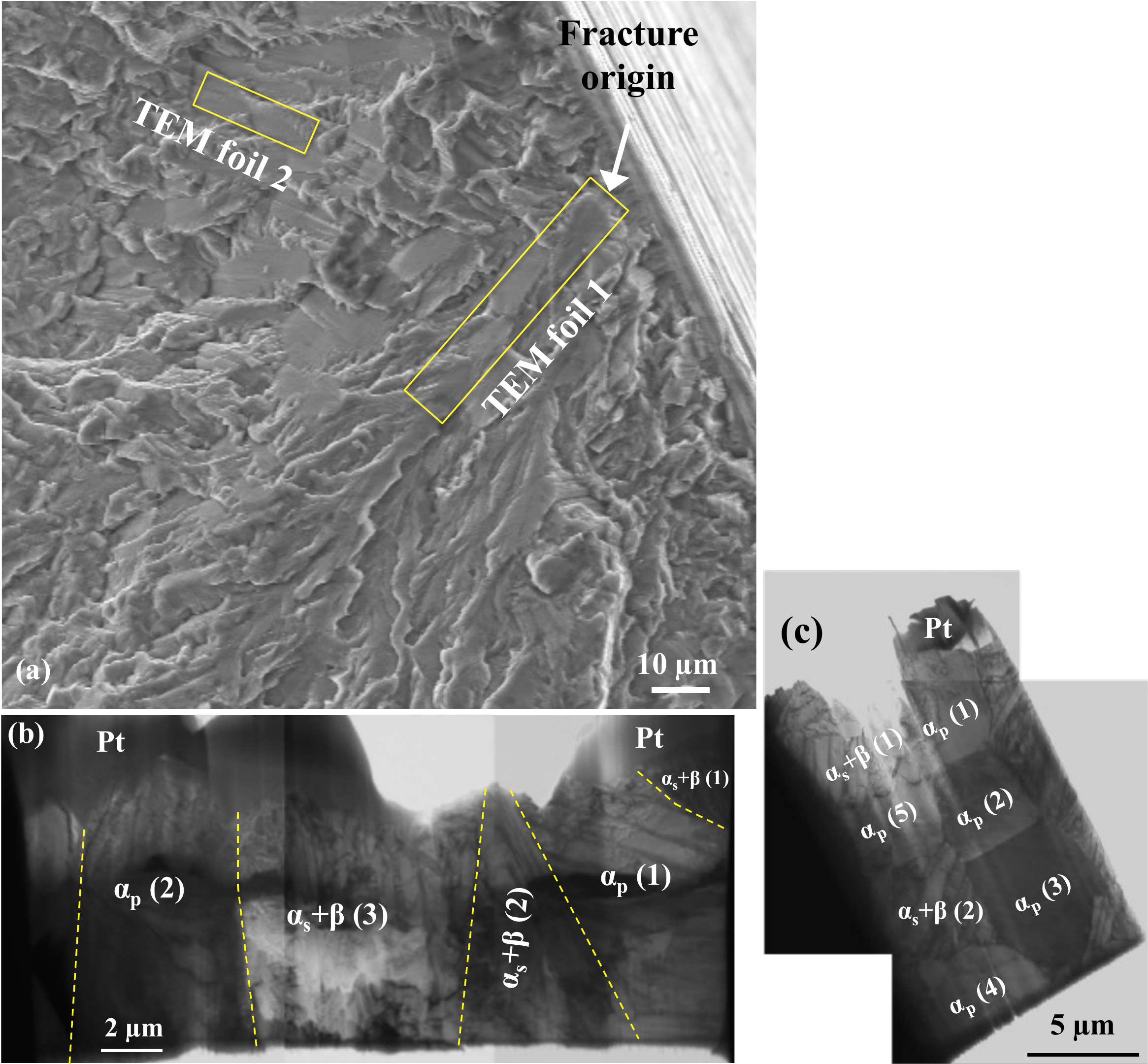}\end{center}
\caption{(a) SEM fractograph showing the regions from which the TEM foils were extracted by FIB Ga ion milling from the fracture origin of low cycle fatigued Ti 6242Si. TEM BF composite micrographs showing (b) foil 1 from the crack origin and (c) foil 2, $50\usk\micro\meter$ away from the crack origin, having different regions marked.}
\end{figure}

\subsection{Dislocation activities at the crack origin}
Foil 1 contained three different two phase regions and two $\alpha_p$ regions, Figure 1(b). The dislocation activities in the two-phase $\alpha_s+\beta$(1) region under a two-beam condition with $B\approx [11\bar{2}3]$ and $g = [0\bar{1}11]$ are shown in Figure 2. The $g.b$ invisibility analysis shows that these are $\langle a\rangle$ type dislocations with Burgers vector $(a/3)[\bar{1}\bar{1}20]$ and $(a/3)[\bar{1}2\bar{1}0]$ in the $\alpha_s$ plates. They are found to be of screw character, gliding on the ($\bar{1}$100) and ($\bar{1}$010) prism planes, respectively. Further, they are gliding on their respective planes without much interaction. The dislocations near the boundary with $\alpha_p(1)$ are shown in Figure~2b.

\begin{table*}[t]\centering
\begin{tabular}{lll}\hline
Region
& Types &Burgers vector and habit plane \\\hline
Primary alpha grain &  & \\
$\alpha_p(1)$ & Slip bands 1-4 & (a/3)$[1\bar{2}13$]($\bar{1}101)$\\
 & Long dislocation lines & (a/3)$[\bar{2}110$]($0001)$\\
 & Random dislocation lines & (a/3)$[\bar{2}110$]\\
$\alpha_p(2)$ & Random dislocation lines & (a/3)$[1\bar{2}13$]\\\hline
Two-phase region &  &  \\
$\alpha_s+\beta(1)$& $a_2$ glide & (a/3)$[\bar{1}2\bar{1}0$]($\bar{1}010)$\\ 
 & $a_3$ glide & (a/3)$[\bar{1}\bar{1}20$]($\bar{1}100)$\\\hline 
\end{tabular}
\caption{Slip systems observed in foil 1.}
\end{table*}
\begin{figure}[b!]
\begin{center}\includegraphics[width=90mm]{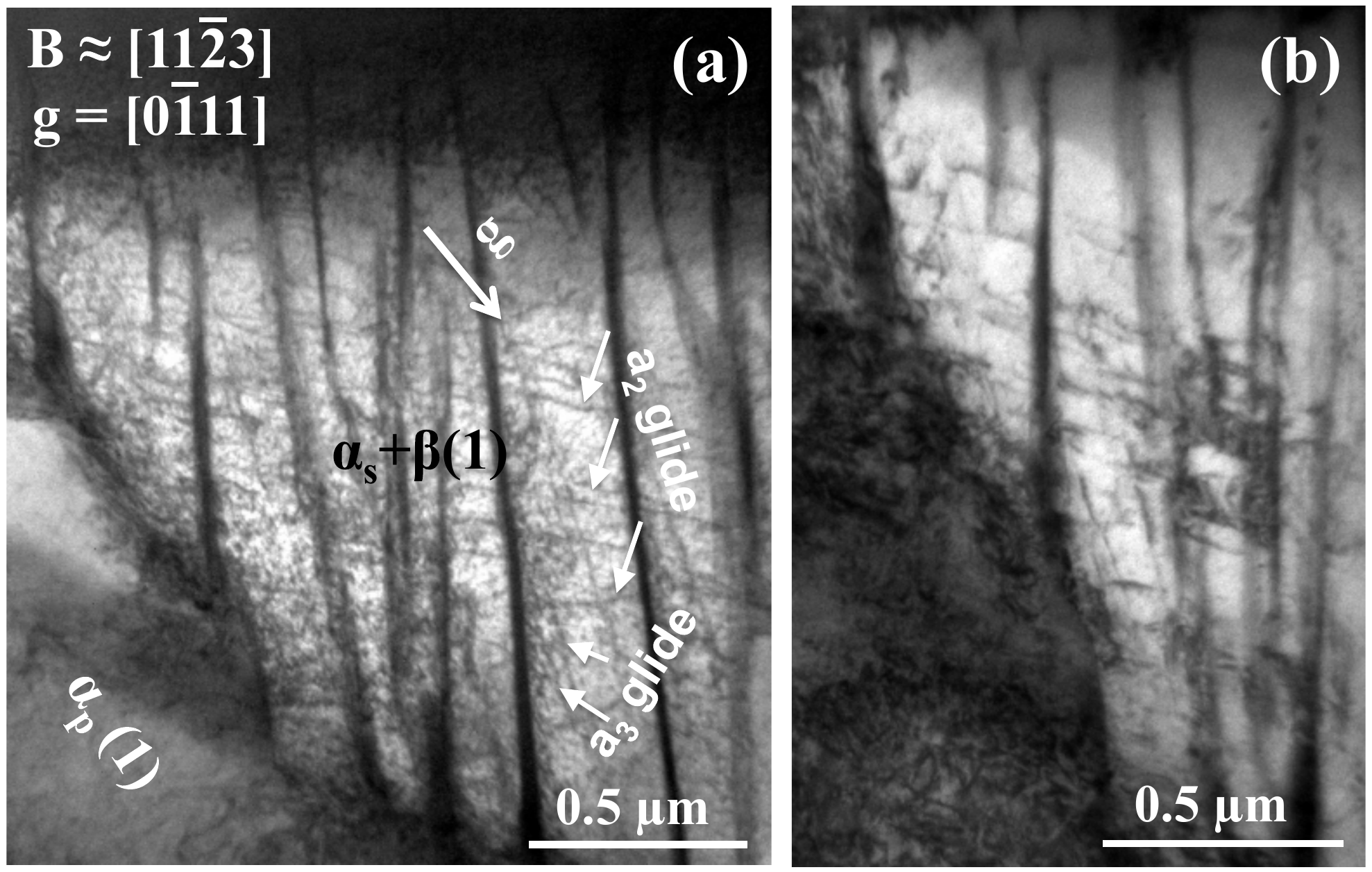}\end{center}
\caption{BF-TEM image showing dislocations in region $\alpha_s+\beta(1)$ in foil 1: (a) overall view and (b) dislocations near the boundary under two beam conditions. The two-beam condition is shown in the inset.}
\end{figure}

A high dislocation density was observed in initiating facet $\alpha_p(1)$, Figure 3. Different illumination conditions are shown in each micrograph; intense planar pile-ups, long straight dislocations and random dislocation lines can be discerned. The dislocation bands (marked SB) are found to be of $\langle c+a \rangle$ type with $b = (a/3)[1\bar{2}13]$, having mixed character and gliding on the first order pyramidal plane ($\bar{1}$101), Figure 3a. The slip bands are around 350--$450\usk\nano\meter$ in thickness, very much higher than classical pile-ups. The dislocations in this array are also dense and tightly packed. Further, each slip band is not a single pile-up but instead consists of a double ended pile-up, Figure 3b. The long straight dislocations indicated by black arrows in Figure 3a are found to be of $\langle a \rangle$ type with Burgers vector $b = (a/3)[\bar{2}110]$ having edge character gliding on the basal plane.  The random dislocation lines in Figure 3c are found to be of $\langle a \rangle$ type with $b = (a/3)[\bar{2}110]$.

\begin{figure*}[t!]
\begin{center}\includegraphics[width=170mm]{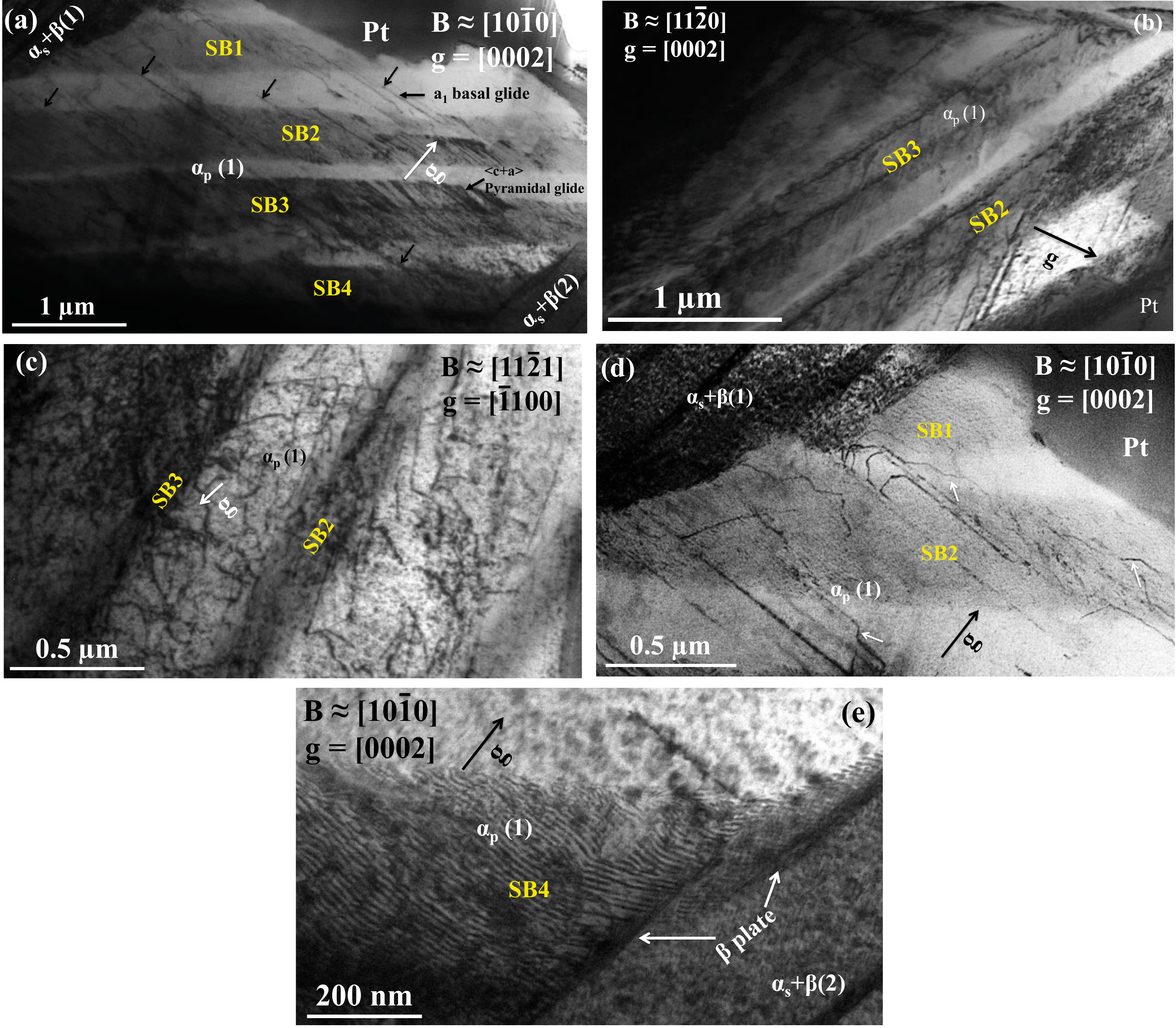}\end{center}
\caption{BF-TEM image showing dislocation activities in grain $\alpha_p(1)$ of foil 1 in the two beam condition. (a) Intense planar pile-ups and long dislocation lines, (b) double ended dislocation pile-up, (c) random dislocation lines in the matrix, (d) jog formation and pile-up of mixed components of $\langle a \rangle$ dislocations near the boundary and (e) shearing of $\beta$ plates by slip band SB4. The two-beam conditions are inset.}
\end{figure*}
	
	To investigate the nature of the dislocations in the pile-ups, the image displacement method \cite{richardson2012systematic} was used. This method is one of the useful dislocation image contrasts which determines whether a pair of dislocations are of same or opposite sign, by viewing them with opposite operating reflections, \emph{i.e.}, $+g$ and $-g$. The dislocations in the double ended pile-ups were captured under two different $g$'s, $[0002]$ and $[000\bar{2}]$ near $B \approx [2\bar{1}\bar{1}0]$, Figure 4.  One dislocation in each pile-up which could be identified without any ambiguity in both the $+g$ and $-g$ conditions are shown by arrows. The separation between the pair of dislocations changes upon changing the sign of $g$, which means that the dislocations are of opposite sign. There will be no change in separation when the dislocations are of same sign; therefore it is concluded that the dislocation on neighbouring slip planes in the double ended pile-ups are of opposite sign.    
    
\begin{figure}[t]
\begin{center}\includegraphics[width=65mm]{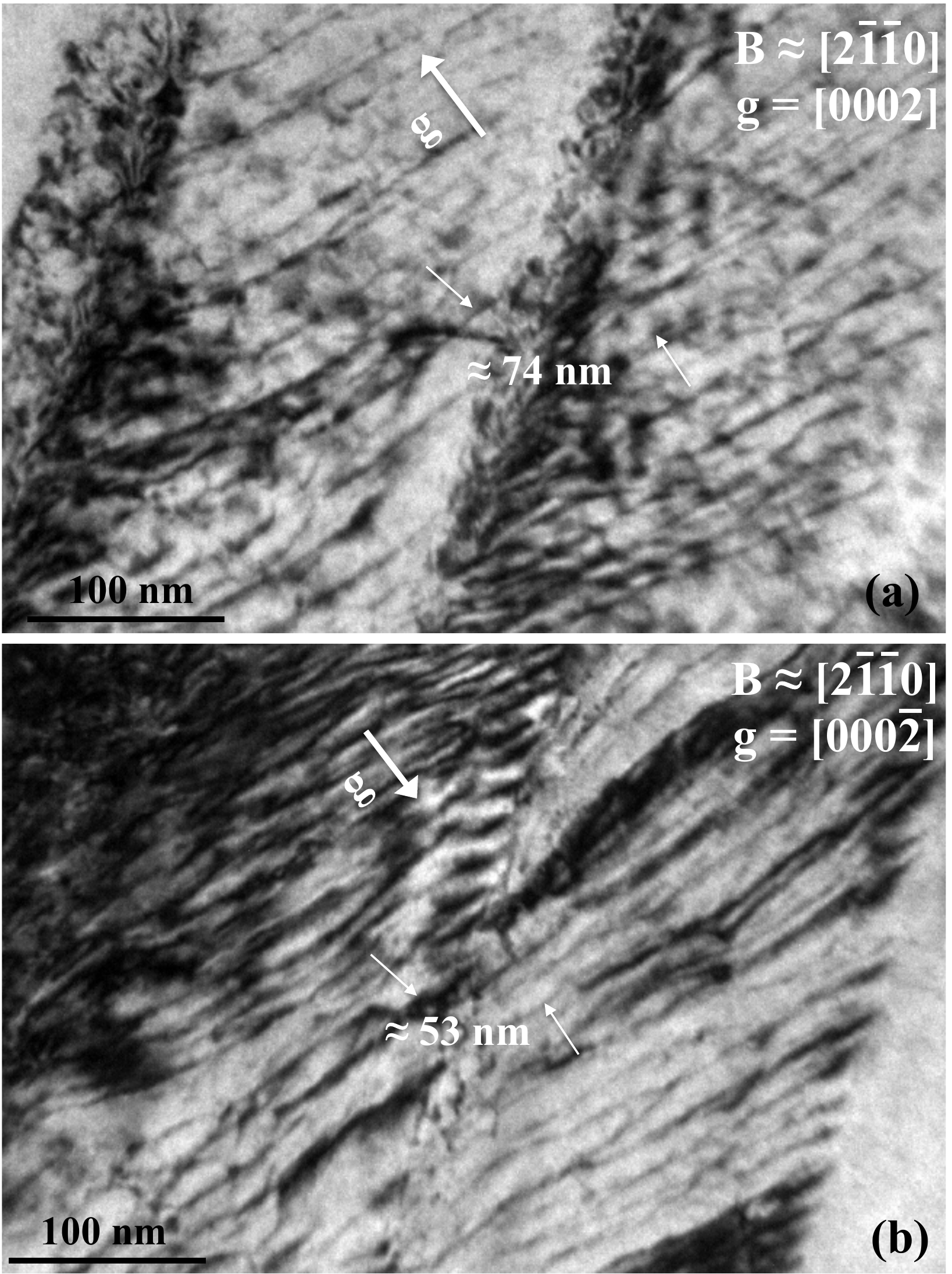}\end{center}
\caption{BF-TEM micrographs in two-beam condition showing the dislocations in slip band SB3 in primary alpha grain $\alpha_p(1)$ of foil 1 when (a) $g = [0002]$ and (b) $g = [000\bar{2}]$ near $B \approx [2\bar{1}\bar{1}0]$. The distance between the dislocations marked are closer in (b) than in (a), showing that the dislocations are of opposite sign.}
\end{figure}
	
	The interactions between the long dislocation lines and the other dislocations resulted in the formation of superjogs in the long dislocation lines, indicated by white arrows in Figure 3d. It can be noted that the mixed component of the long dislocation lines are found to pile-up in the boundary between $\alpha_s + \beta$(1) and $\alpha_p(1)$, Figure 3d. It is also observed that the slip bands in this grain did not transfer across the boundary between the $\alpha_p$ grain and the two phase region, even though they are intense and large. However, the slip bands were observed to shear the $\beta$ plate in the $\alpha_s + \beta$ (3) region, Figure 3e. 
	
	Figure 5 shows the dislocations observed in $\alpha_p(2)$, under the two beam condition with $g = [0002]$ near $B\approx [01\bar{1}0]$. These dislocations are of $\langle c+a \rangle$ type with $b = (a/3)[1\bar{2}13]$, and are not observed to pile up on a particular plane. Similar to region $\alpha_s$ + $\beta$ (1), $\alpha_s$ + $\beta$  (2) also showed prismatic glide of $\langle a \rangle$ type dislocations. Relatively few dislocations were observed in region $\alpha_s$ + $\beta$ (3). For clarity, the dislocations in regions $\alpha_s$ + $\beta$ (2) and $\alpha_s$ + $\beta$ (3) will not be discussed further, but similar findings were obtained. \textcolor{black}{The slip systems observed in foil 1 are summarised in Table 1.}
	
\begin{figure}[t]
\begin{center}\includegraphics[width=50mm]{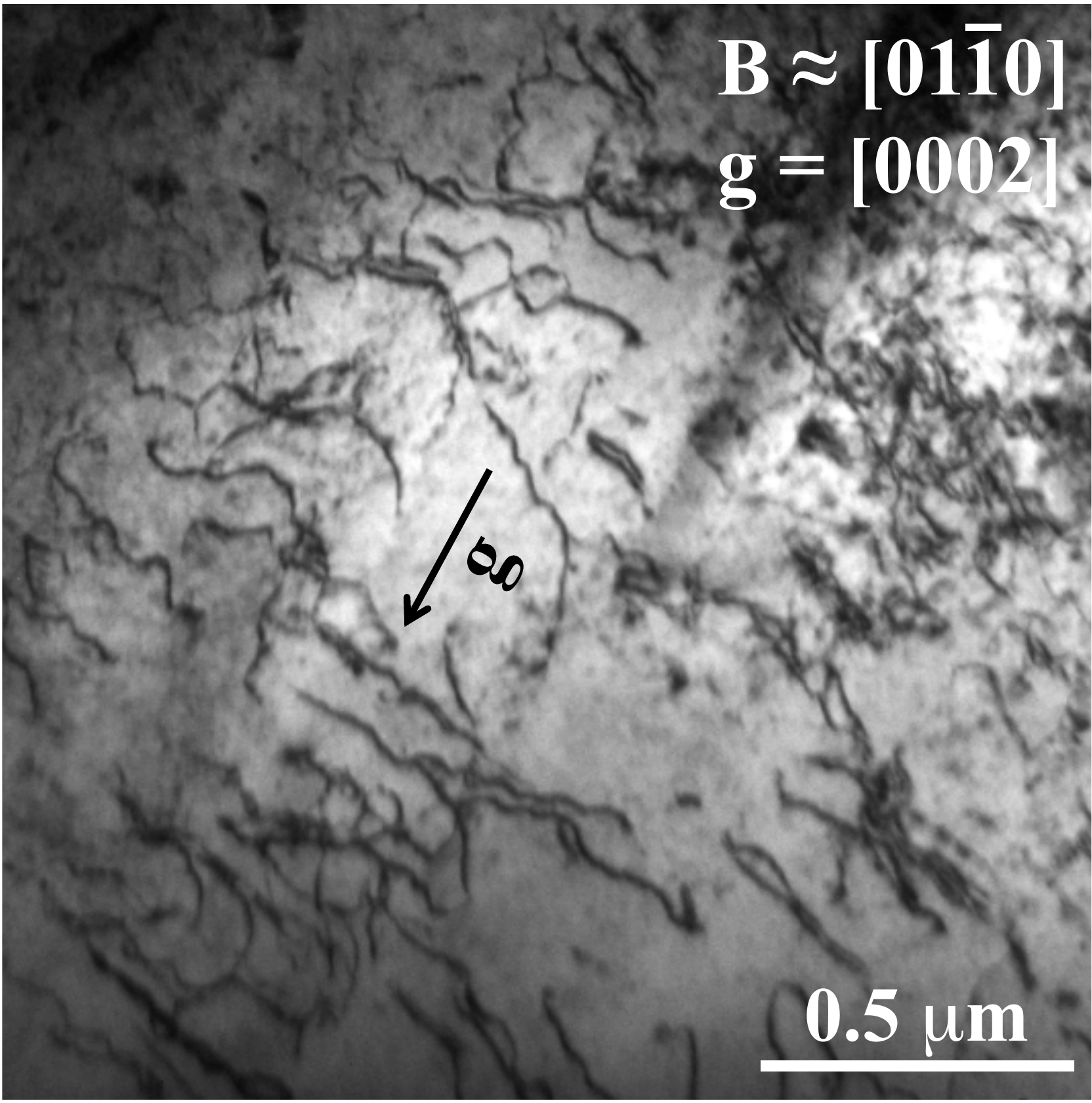}\end{center}
\caption{BF-TEM image showing $\langle c+a\rangle$ dislocations in $\alpha_p(2)$ grain of foil 1 in the two-beam condition with $g = [0002]$ near $B \approx [01\bar{1}0]$.}
\end{figure} 

\subsection{Dislocation activities during fatigue crack growth, $50\usk\micro\meter$ away from the nucleation site}      
This section discusses the dislocation activities in foil 2, Figure 1(c), which consists of many similarly oriented primary alpha grains and some two-phase regions.
    
\subsubsection{Slip transfer across the $\alpha_p$ grains}
The slip activities across the five different $\alpha_p$ grains in foil 2 are shown in Figure 6, which is a composite micrograph of bright field images taken when each grain is tilted to its two-beam condition individually. The crystallographic orientation of these grains normal to the loading direction was obtained by TKD of the TEM foil, Table 2. Grains 1 to 4 possessed almost similar inclinations of their $c$-axes to the loading direction, with only grain $\alpha_p(5)$ having its $c$-axis $<45\degree$ from the loading direction. 

\begin{table}[b]\centering
\begin{tabular}{cc}\hline
Grain & Orientation of\\
& $c$-axis (degrees)\\\hline
1 & 47.6\\
2 & 52.4\\
3 & 52.4 \\
4 & 58.5\\
5 & 32.0\\\hline
\end{tabular}
\caption{Crystallographic orientation of the $\alpha_p$ grains with the loading direction in foil 2.}
\end{table}

\begin{figure}[t!]
\begin{center}\includegraphics[width=90mm]{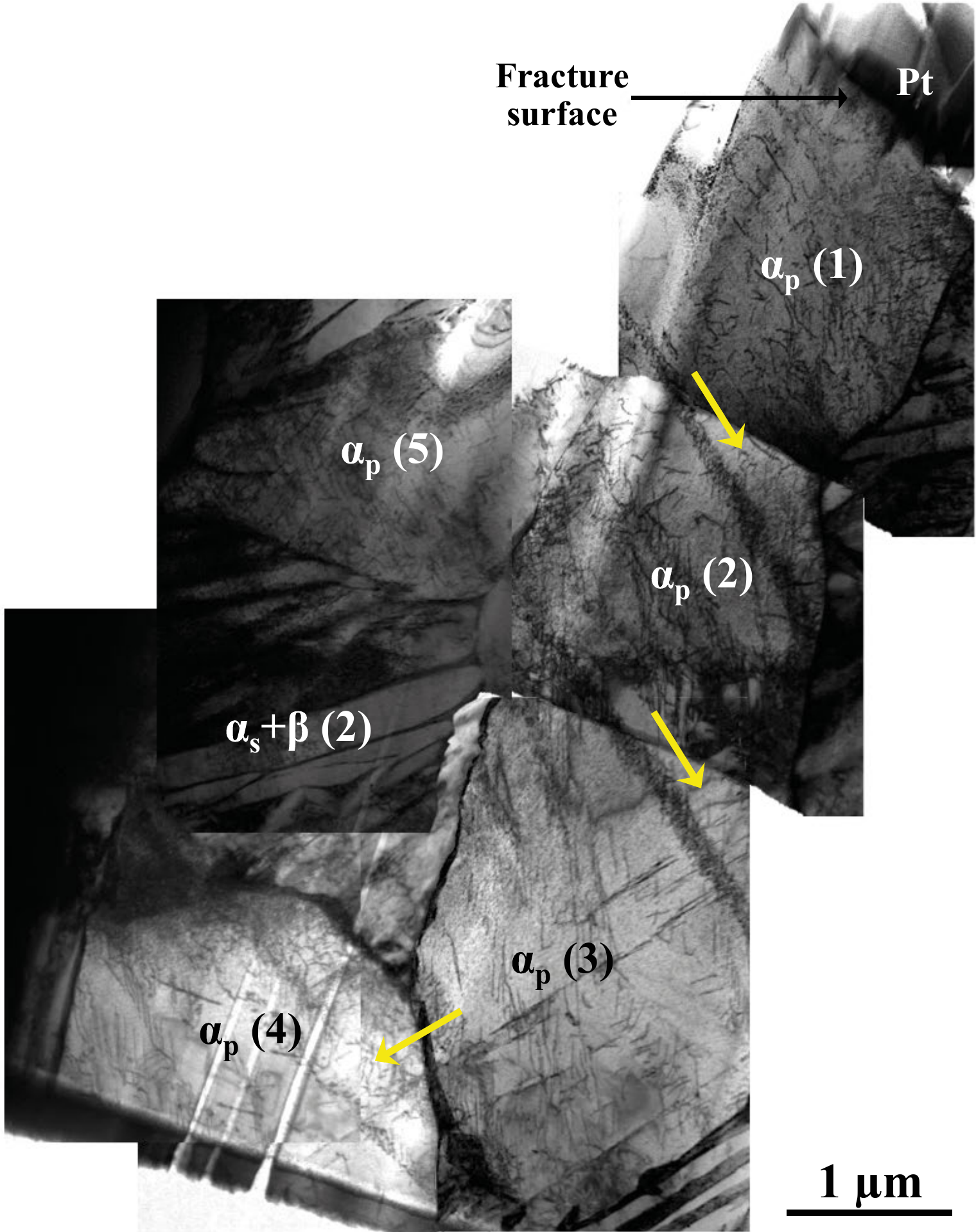}\end{center}
\caption{BF-TEM composite micrograph showing the slip transfer across the primary alpha grains in foil 2. Each grain has been tilted individually to the two-beam condition to provide diffraction contrast of the dislocations. Slip transfer across the grains is highlighted by arrows.}
\end{figure} 

It was observed that the dislocations in the first four grains (1-4) piled-up, forming well-defined slip bands, and that those slip bands underwent slip transfer across the $\alpha_p$ grain boundaries (indicated by arrows in Figure 6). Between grains 1/2 and 2/3, the piled up dislocations transferred across the boundary whereas between grains 3/4, just a single dislocation line was observed to transfer.  In contrast, grain 5 shows neither pile up of dislocations nor slip transfer.
	
	The slip activity in  primary alpha grain $\alpha_p$ (1) is shown in Figure 7, captured under a two-beam condition of $B \approx [2\bar{1}\bar{1}3]$ and $g=[10\bar{1}\bar{1}]$. The dislocations mainly piled-up on a particular slip plane and some dislocations are also observed between these pile-ups. The dislocations in the pile-up are of $(a/3)[\bar{2}110]$ type and gliding on basal planes, marked as slip bands SB1, SB2 and SB3. The pile-up in SB1 was observed to be double ended. In addition to these dislocations on basal planes, the other $(a/3)[11\bar{2}0]$ type dislocations are found to glide in a random way, Figure 7b. Of the three slip bands observed in this grain, SB3 was found to transfer across the boundary, forming SB4 in primary alpha grain $\alpha_p$(2) . 
	
\begin{figure}[t]
\begin{center}\includegraphics[width=90mm]{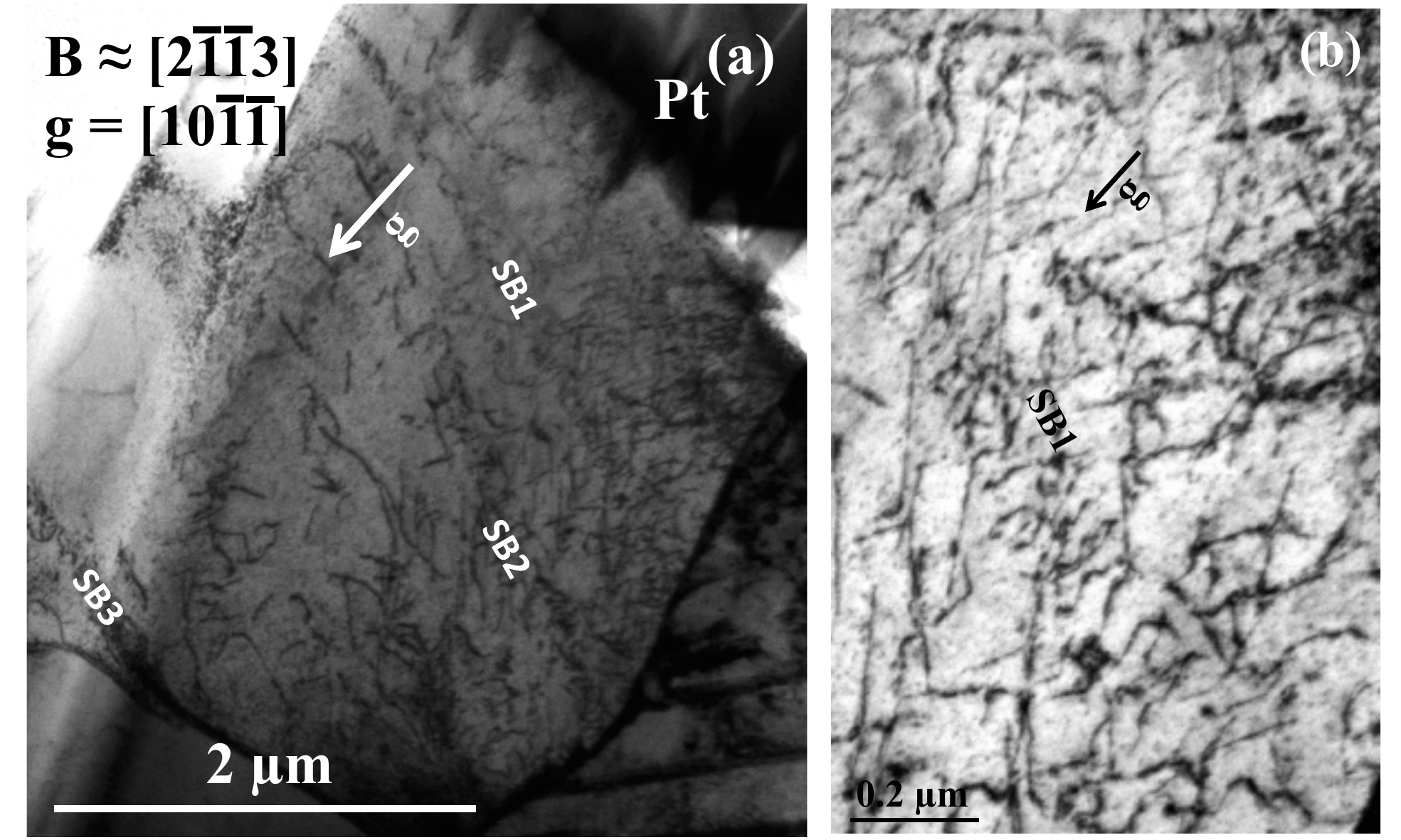}\end{center}
\caption{BF-TEM image showing slip activities in primary alpha grain $\alpha_p(1)$ of foil 2. (a) overall view and (b) high magnification image showing a pile-up (SB1) and other random dislocations. The two-beam condition is shown in the inset of the figure. }
\end{figure}
    
	Figure 8 shows a high dislocation density in grain $\alpha_p$(2), mainly as pile-ups (SB4-SB8), with some random dislocations. The dislocations in this grain are shown under two different two beam conditions with $B \approx [2\bar{1}\bar{1}3]$ and $B \approx [4\bar{2}\bar{2}3]$. This grain shares its boundary with grains 1, 3 and 5 and the two-phase region $\alpha_s + \beta$(2). All the slip bands observed in this grain were of $(a/3)[\bar{2}110]$ basal slip character, as observed in grain 1. SB8 was found to transfer across the boundary, forming SB9 in grain 3. Slip transfer was not observed from grain 2 to grain 5 and the neighbouring two-phase region. In addition, some random dislocation lines were also observed, of $(a/3)[11\bar{2}0]$ type, Figure 8b.
	
\begin{figure}[t]
\begin{center}\includegraphics[width=90mm]{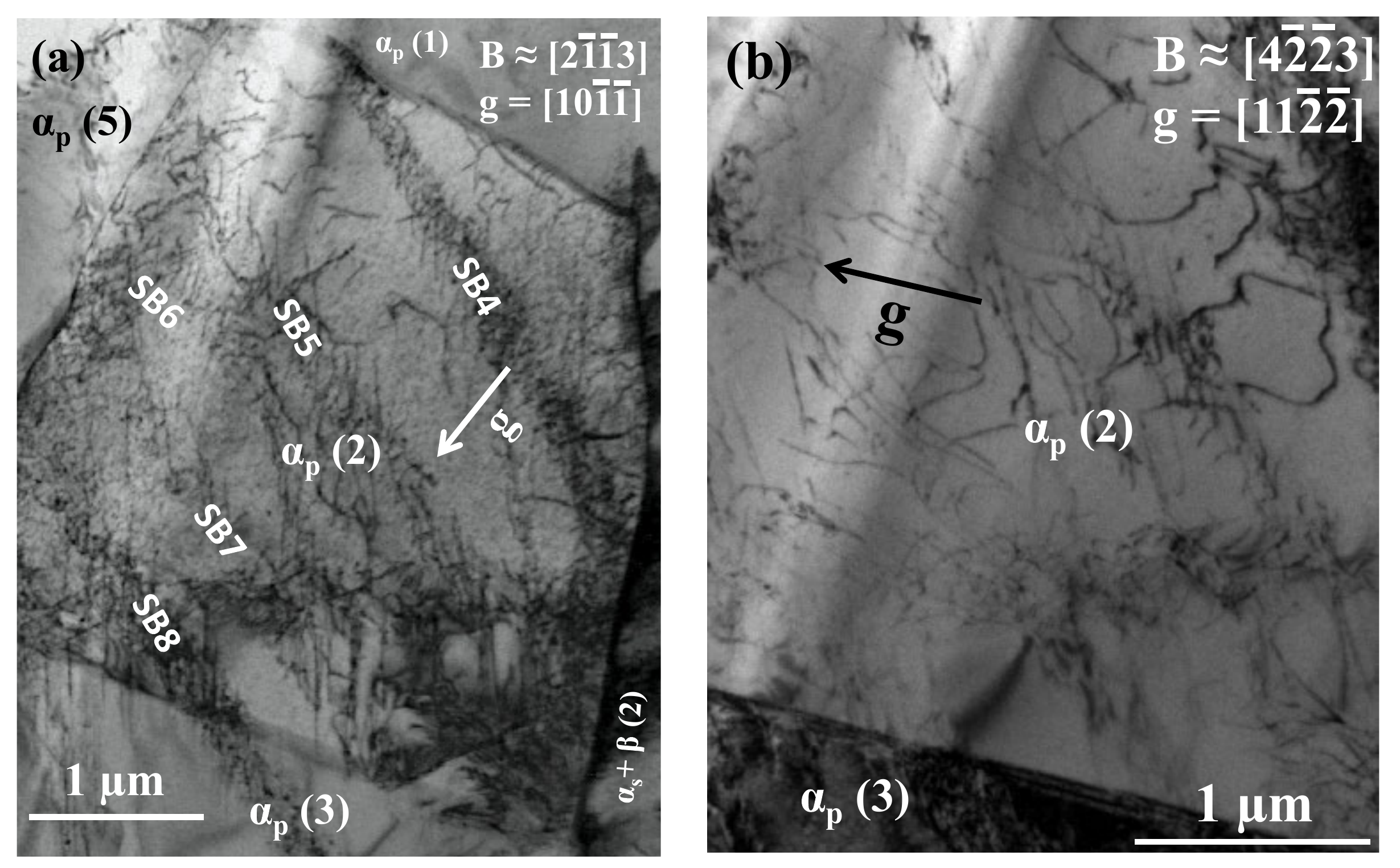}\end{center}
\caption{BF-TEM image showing slip activities in primary alpha grain $\alpha_p(2)$ of foil 2 under two-beam conditions when (a) $B \approx [2\bar{1}\bar{1}3]$ and (b) $B \approx [4\bar{2}\bar{2}3]$. The two-beam conditions are inset.}
\end{figure}
    
	The slip activity in $\alpha_p$(3) is shown in Figure 9, which is a bright field image captured in the two-beam condition with $B \approx [10\bar{1}0]$  and $g = [\bar{1}2\bar{1}0]$.  The various slip bands are marked. Slip bands 9, 10, 12 and 14 are $(a/3)[\bar{2}110]$ basal and slip bands 11 and 13 are $(a/3)[1\bar{2}10]$  basal. Slip band 9 traversed the whole grain,  indicating that it was the primarily activated deformation mode, while the other slip bands will be secondary slip activated after reaching a critical stress. There are other long dislocation lines in addition to the slip bands, of $(a/3)[\bar{1}\bar{1}20]$ and $(a/3)[\bar{1}2\bar{1}0]$ type (dotted arrows). These dislocations were found to transfer the boundary between grains 3 and 4, where a thin $\beta$ ligament was observed between the grains, Figure 6. Thus this grain showed all the three types of $\langle a \rangle$ dislocations, resulting in hexagonal networks of dislocations.
    
\begin{figure}[t]
\begin{center}\includegraphics[width=50mm]{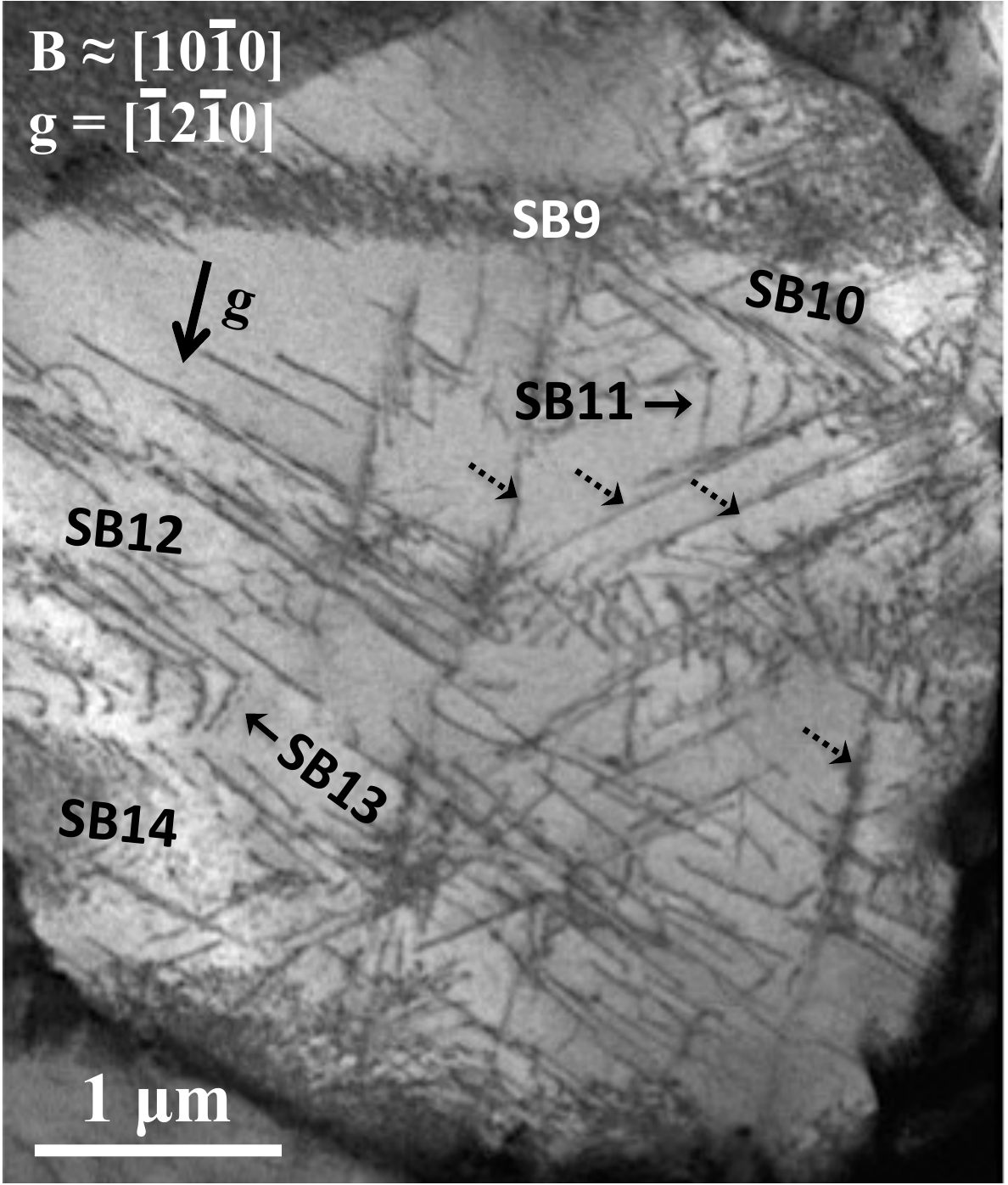}\end{center}
\caption{BF-TEM image showing the dislocation configurations in primary alpha grain $\alpha_p(3)$ of foil 2, under two-beam conditions near $B \approx [10\bar{1}0]$.}
\end{figure}
	
	Figure 10 shows the slip activity in primary alpha grain $\alpha_p$(4), which was similarly found to deform mainly by planar slip. All the slip bands are $(a/3)[\bar{2}110]$ basal except SB17 which was found to be of $(a/3)[1\bar{2}10]$ basal character. Many dislocation loops were observed near the boundary between this grain and the two-phase region $\alpha_s + \beta$(2), Figure 10b. Dislocation dipoles were also observed in this grain, shown by dotted arrows in Figure 10c. These dipoles are expected to form by encounters between dislocations of opposite sign. They are observed to be narrow screw dislocation dipoles, which are unstable and would be expected to decompose into invisible debris of near-atomic dimension. 

\begin{figure*}[t]
\begin{center}\includegraphics[width=150mm]{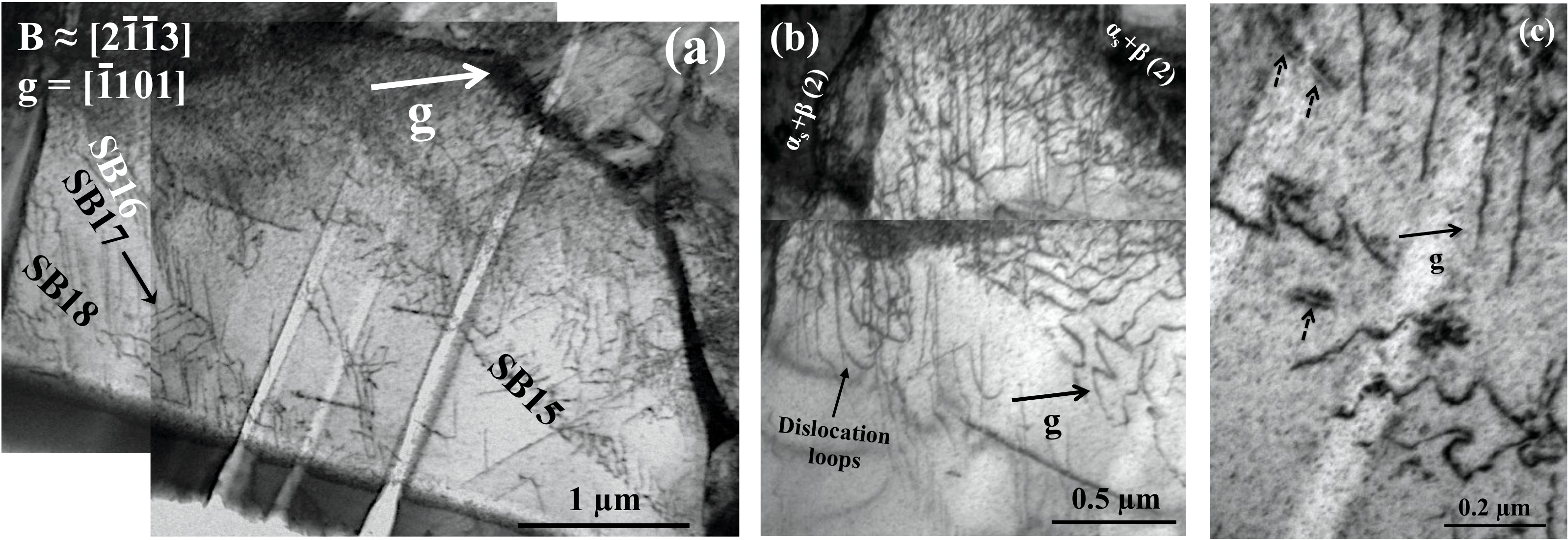}\end{center}
\caption{BF-TEM image showing dislocations in primary alpha grain $\alpha_p(4)$ of foil 2 under two-beam conditions: (a) overall view, (b) dislocation loops from the $\alpha_p/(\alpha_s+\beta)$ boundary, and (c) dislocation dipoles.  (a) and (b) are composite micrographs and the two-beam condition is inset.}
\end{figure*}
	
	In contrast, grain 5 did not show planar slip bands; instead homogeneous deformation was observed, Figure 11. Both $\langle a \rangle$ and $\langle c+a \rangle$ type dislocations were observed in this grain. The $\langle a \rangle$  dislocations were of $(a/3)[\bar{2}110]$ and $(a/3)[11\bar{2}0]$ type and were not observed on particular slip plane, Figure 11a.  The $\langle c+a \rangle$  dislocations were of $(a/3)[\bar{2}113]$ type and were few in number and density, Figure 11b. \textcolor{black}{The slip systems observed in the primary alpha grains of foil 2 are listed in Table 3}

\begin{figure}[b!]
\begin{center}\includegraphics[width=90mm]{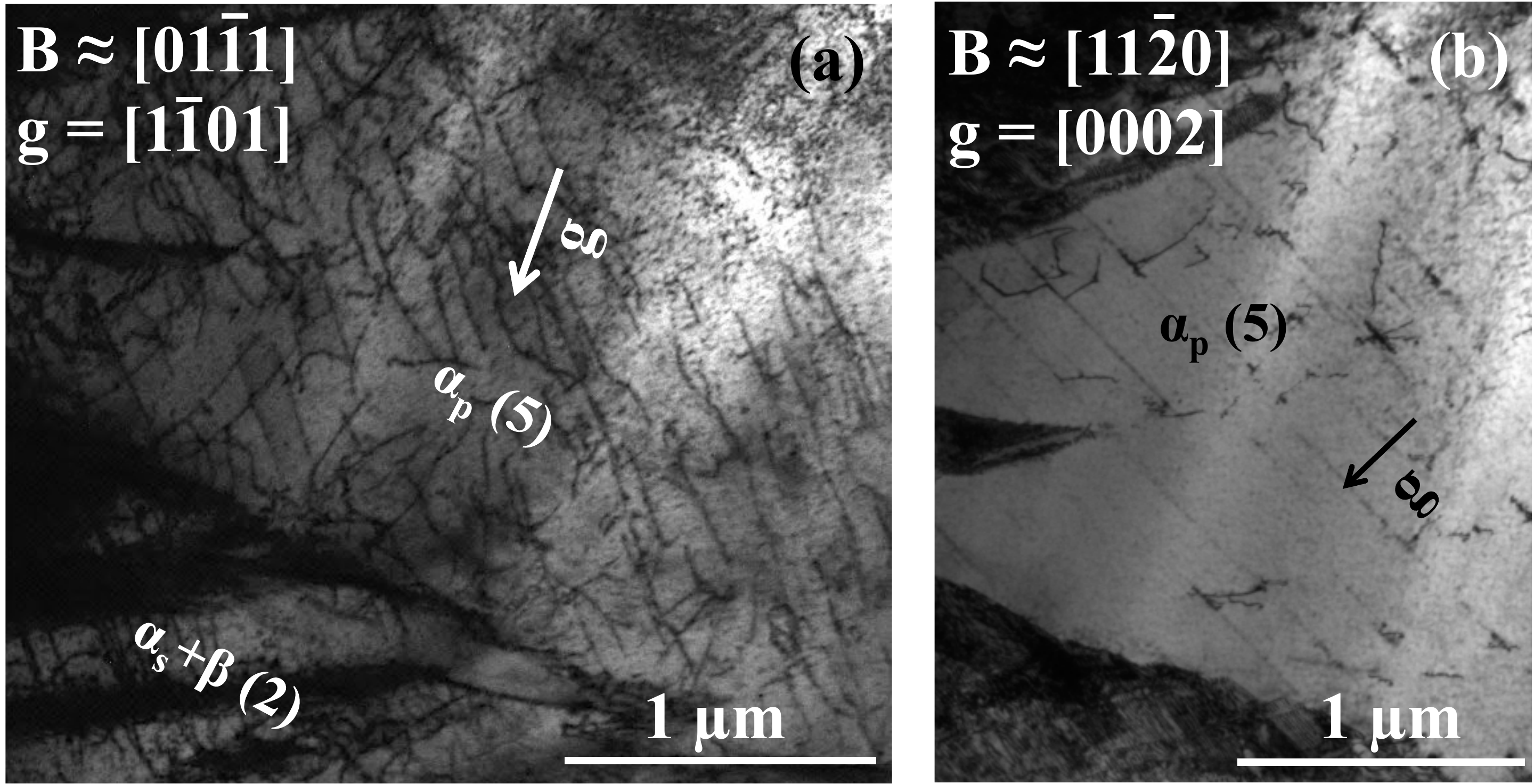}\end{center}
\caption{BF-TEM image showing dislocations in primary alpha grain $\alpha_p(5)$ of foil 2 in  two-beam conditions: (a) overall view showing $\langle a \rangle$ type dislocations under $B \approx [01\bar{1}1]$ and (b) $\langle c+a \rangle$ dislocations under $B \approx [11\bar{2}0]$. The two-beam condition is shown in the inset.}
\end{figure}

	Figure 12 shows the various types of dislocation interactions near the boundaries (inclined in the foil; shadowed) of the primary alpha grains in foil 2. The interaction is mostly in the form of slip transfer of piled-up dislocations, Figure 12a, with some are slip transfer of dislocation groups, \emph{e.g.} Figure 12b. At the grain 2/3 boundary, the impingement of  dislocations has caused nucleation of dislocations in the same impinging grain 2, Figure 12c. This might be due to the generation of dislocation sources in the grain boundary by the impinging dislocations. Alternately, grain boundary blocking of dislocations can be observed in Figure 12d.
	
	\begin{table*}[t]\centering\small
\begin{tabular}{lll}\hline
Region
& Types &Burgers vector and habit plane \\\hline
Primary alpha grain &  & \\
$\alpha_p(1)$ & Slip bands 1-3 & (a/3)$[\bar{2}110]$basal\\
  & Random dislocation lines & (a/3)$[11\bar{2}0$]\\
$\alpha_p(2)$ & Slip bands 4-8 & (a/3)$[\bar{2}110]$basal\\
 &Random dislocation lines&(a/3)$[11\bar{2}0]$basal\\
$\alpha_p(3)$ & Slip bands 9,10,12,14 & (a/3)$[\bar{2}110]$basal\\
 & Slip bands 11,13 & (a/3)$[1\bar{2}10]$basal\\ 
  & Long dislocation lines & (a/3)$[\bar{1}\bar{1}20]$ and (a/3)$[\bar{1}2\bar{1}0]$ \\ 
  $\alpha_p(4)$ & Slip bands 15,16,18&(a/3)$[\bar{2}110]$basal\\
  &Slip band 17&(a/3)$[1\bar{2}10]$basal\\
$\alpha_p(5)$ & Random dislocation lines&(a/3)$[\bar{2}110]$,(a/3)$[11\bar{2}0]$ and (a/3)$[\bar{2}113]$\\\hline
Two-phase region & & \\
$\alpha_{s1}$&Dislocation loops&(a/3)$[11\bar{2}0]$basal\\
 &Straight dislocations&(a/3)$[\bar{2}11\bar{3}]$pyramidal\\
$\alpha_{s2}$&Dislocation loops&(a/3)$[11\bar{2}0]$basal\\
&Pile-up&(a/3)$[\bar{2}110]$\\
$\alpha_{s3}$&Dislocation loops&(a/3)$[11\bar{2}0]$basal\\
&Dislocation network&(a/3)$[11\bar{2}0]$\\
$\beta_2$&Dislocation network&(a/2)$[11\bar{1}]\{211\}$ and (a/2)$[\bar{1}11]$\{211\} \\
&Straight dislocations&(a/2)$[010]$\\\hline
\end{tabular}
\caption{Slip systems observed in foil 2.}
\end{table*}

\begin{figure*}[t!]
\begin{center}\includegraphics[width=150mm]{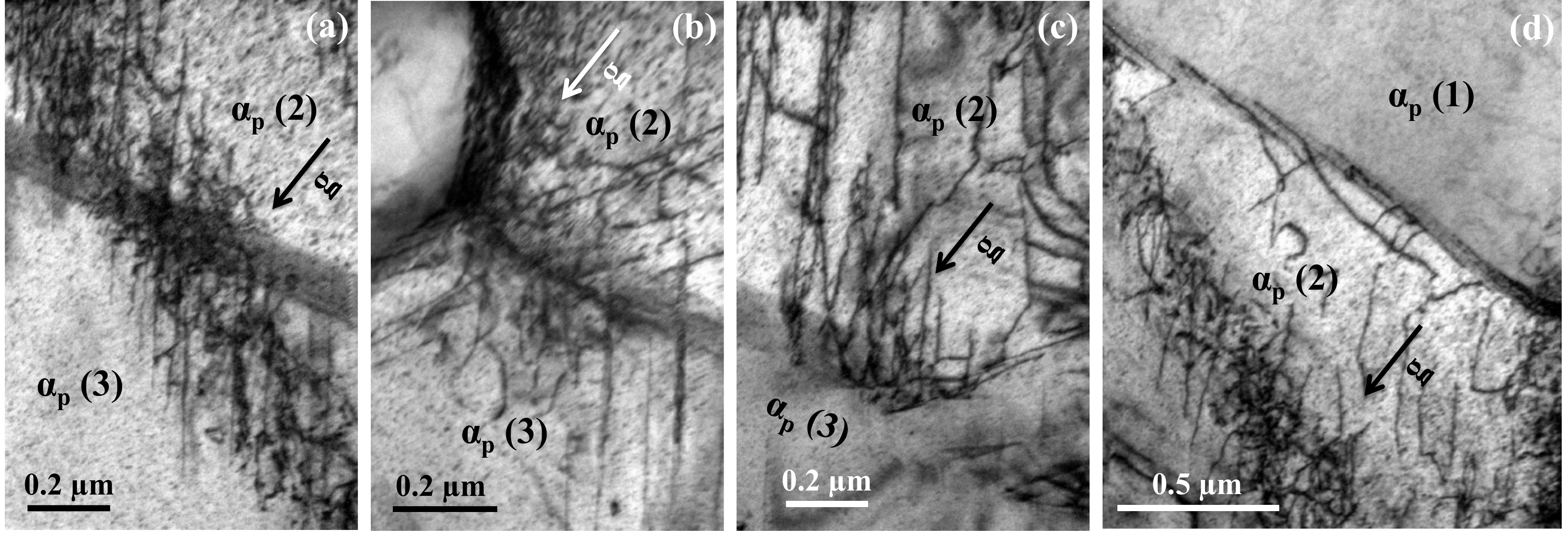}\end{center}
\caption{BF-TEM image showing dislocation interactions at the boundary between: (a-c) primary alpha grains $\alpha_p(2)/ \alpha_p(3)$ and (d) primary alpha grains $\alpha_p(1)/ \alpha_p(2)$ when grain 2 is tilted to $g = [10\bar{1}\bar{1}]$  near $B \approx [2\bar{1}\bar{1}3]$. The $g$ direction is shown in the Figure.}
\end{figure*}
    
	\subsubsection{Dislocation activities in two-phase region}
	Two phase region $\alpha_s$ + $\beta$(2) in foil 2 was selected for further study. The various $\alpha_s$ plates and $\beta$ ligaments investigated are shown in Figure 13a, which is the pattern quality map obtained by TKD. The phase map and the orientation map are also shown in Figures 13b-c. The orientation relation between secondary alpha plates $\alpha_{s1}$, $\alpha_{s2}$ and $\beta_1$ is shown in Figure 13d, which is the pole figure obtained from this two-phase region. The $(0001)$ plane normal and the three $\langle a \rangle$ directions of $\alpha_{s1}$ and $\alpha_{s2}$ plates are marked in the $\alpha_s$ pole figures. The \{110\} plane normal and the $\langle b_1 \rangle$ and $\langle b_2 \rangle$ directions are marked in the $\beta_1$ pole figure. The procedure for identifying these plane normals and directions is discussed in~\cite{tong2017using}. It is found that the expected $(101)_\beta || (0001)_\alpha$ and $[1\bar{1}\bar{1}]_\beta || [2\bar{1}\bar{1}0]_\alpha$ Burgers' orientation relation is obeyed~\cite{burgers1934process}.       
    
\begin{figure*}[t!]
\begin{center}\includegraphics[width=150mm]{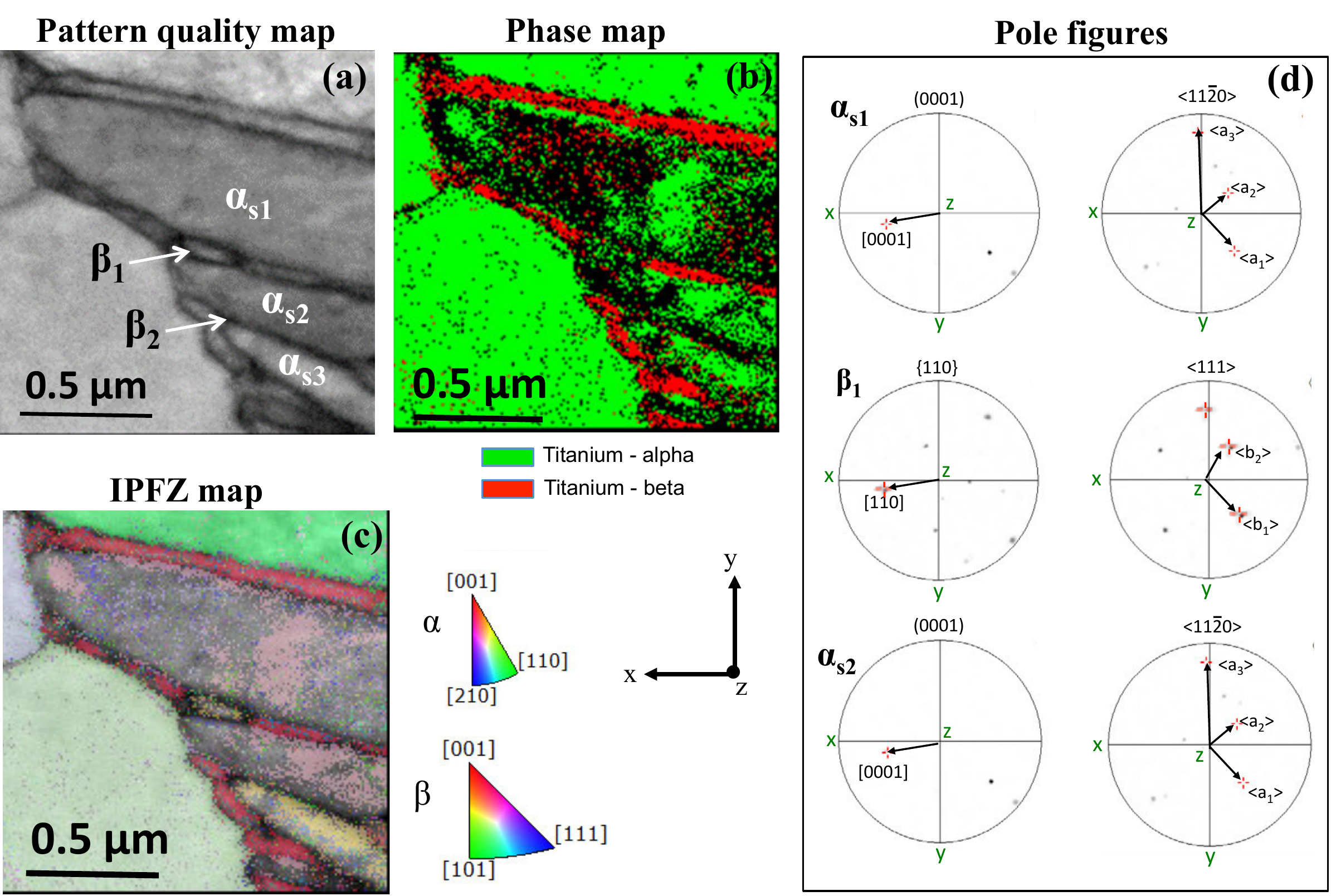}\end{center}
\caption{TKD map in two phase region $\alpha_s+\beta(2)$ of foil 2 showing the orientation relation between the $\alpha_s$ and $\beta$: (a) pattern quality map, (b) phase map, (c) IPFZ map and (d) pole figure showing the orientation of secondary alpha plates $\alpha_{s1}$ $\alpha_{s2}$ and beta plate $\beta_1$.}
\end{figure*} 

	 Figure 14 shows the slip activities in the $\alpha_s$ plates, imaged in the two beam condition using the beam directions $B$ shown. Figures 14(a-b) are bright field images taken with $\alpha_{s2}$ near $B \approx [11\bar{2}0]$ and $g = [000\bar{2}]$; $\alpha_{s3}$ was also in the diffraction condition due to its similar orientation to $\alpha_{s2}$. The $g.b$ analysis shows that the straight dislocations in Figure 14a are of $\langle a \rangle$ type with $b =(a/3)[11\bar{2}0]$, having edge character and gliding on basal planes. The same type of dislocations can also be seen in plate $\alpha_{s1}$, but this plate is slightly misoriented and is therefore darker in contrast. It can be observed that there is a one-to-one correspondence of these dislocations in these $\alpha_s$ plates. A number of dislocation loops are observed to nucleate from the $\alpha_s/\beta$ boundary, Figure 14b. These are found to be $[11\bar{2}0]$-type loops, gliding on basal planes. The nucleation of numerous $(a/3)[11\bar{2}0]$-type dislocations in plate $\alpha_{s2}$ plate was observed in Figure 14c. 
     
\begin{figure*}[t!]
\begin{center}\includegraphics[width=150mm]{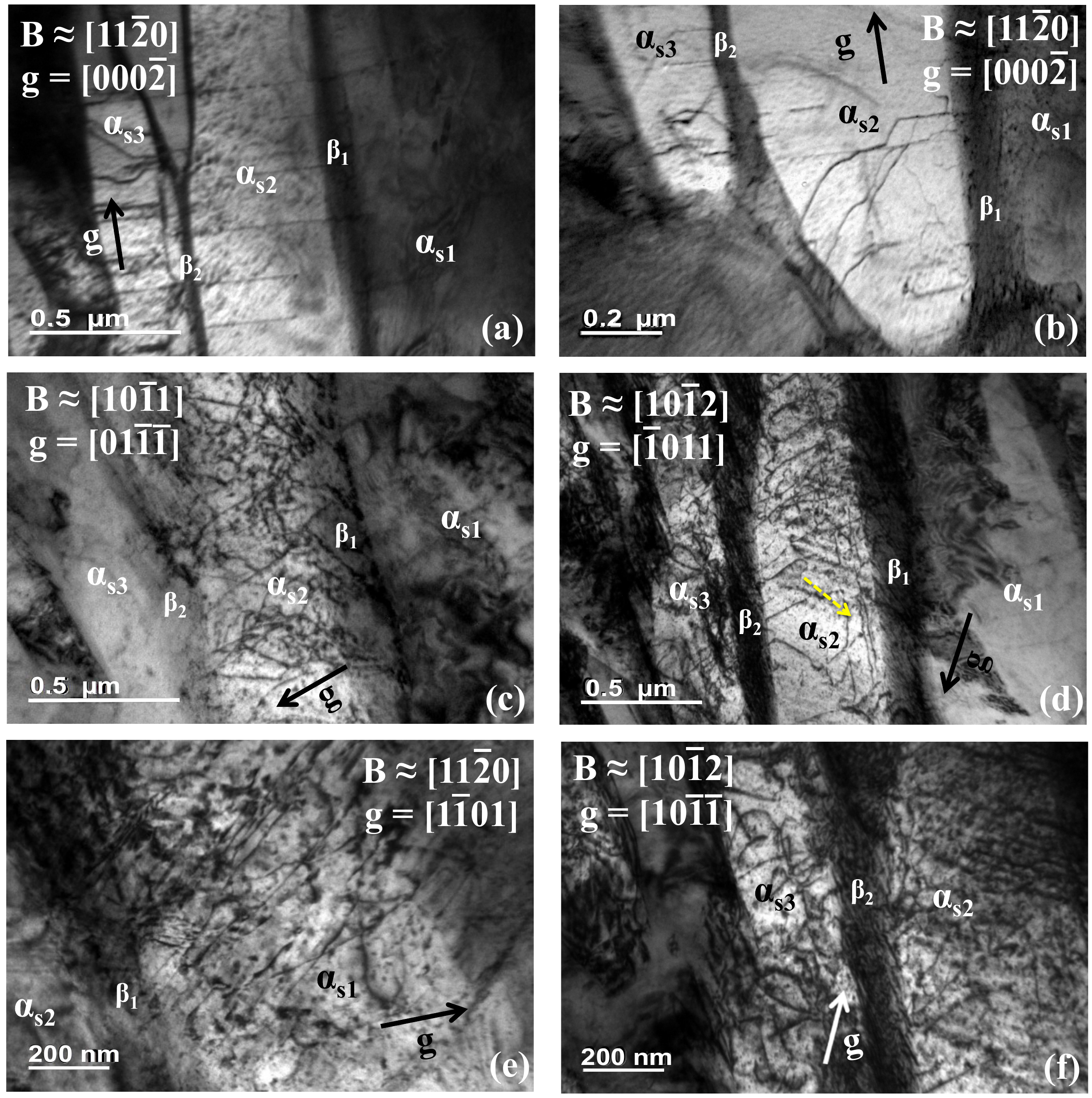}\end{center}
\caption{BF-TEM in two-beam conditions under different beam directions showing dislocations in the $\alpha_s$ plates of two phase region $\alpha_s+\beta(2)$ in foil 2:  (a) one-to-one correspondence of $a_3$ dislocations within the $\alpha$ plates, (b) generation of dislocation loops from the $\alpha_s+\beta$ boundary, (c) nucleation of several $a_3$ dislocations from the $\alpha_s+\beta$ boundary, (d) dislocation pile-up (shown by arrows) near the $\alpha_s+\beta$ boundary, (e) $\langle c+a \rangle$ pyramidal glide in plate $\alpha_{s1}$ and (f) dislocation networks in plate $\alpha_{s3}$.  The two-beam conditions and $g$ directions are shown.}
\end{figure*} 
     
     Figure 14d shows these $(a/3)[11\bar{2}0]$-type dislocations and a pile-up of a small number of $(a/3)[\bar{2}110]$ type dislocations near the $\alpha_{s2}/\beta_1$ boundary, which is shown by a dotted yellow arrow. In addition to these $a_3$-type dislocations, some $\langle c+a \rangle$ dislocations were observed in plate $\alpha_{s1}$, Figure 14e, and a network of $(a/3)[11\bar{2}0]$-type dislocations were observed in plate $\alpha_{s3}$, Figure 14f. The $\langle c+a \rangle$ dislocations in $\alpha_{s1}$ are of $(a/3)[\bar{2}11\bar{3}]$ type, gliding on a first order pyramidal plane.
	 
	 Both retained $\beta$ ligaments $\beta_1$ and $\beta_2$ show similar dislocation activities, so only ligament $\beta_2$ is shown here. Figure 15 shows bright field images under two beam conditions with $B \approx [001]$ and $g$-vectors $[\bar{2}00]$ and $[\bar{1}10]$. A network of dislocations are observed in this $\beta$ plate under $g = [\bar{2}00]$, Figure  15a. Detailed $g.b$ analysis shows that these are $b_1$ $(a/2)[11\bar{1}]$ and $b_2$ $(a/2)[\bar{1}11]$ dislocations gliding  on $\{211\}$-type planes. In addition to these $b_1$ and $b_2$ dislocations, some $b_{010}$ (a/2)[010] dislocations were also observed with $g = [\bar{1}10]$, Figure 15b. It is rare to observe this type of $b_{010}$ dislocation in Ti alloys. \textcolor{black}{The slip systems observed in this two-phase region are summarised in Table 3.}   
     
\begin{figure}[t]
\begin{center}\includegraphics[width=70mm]{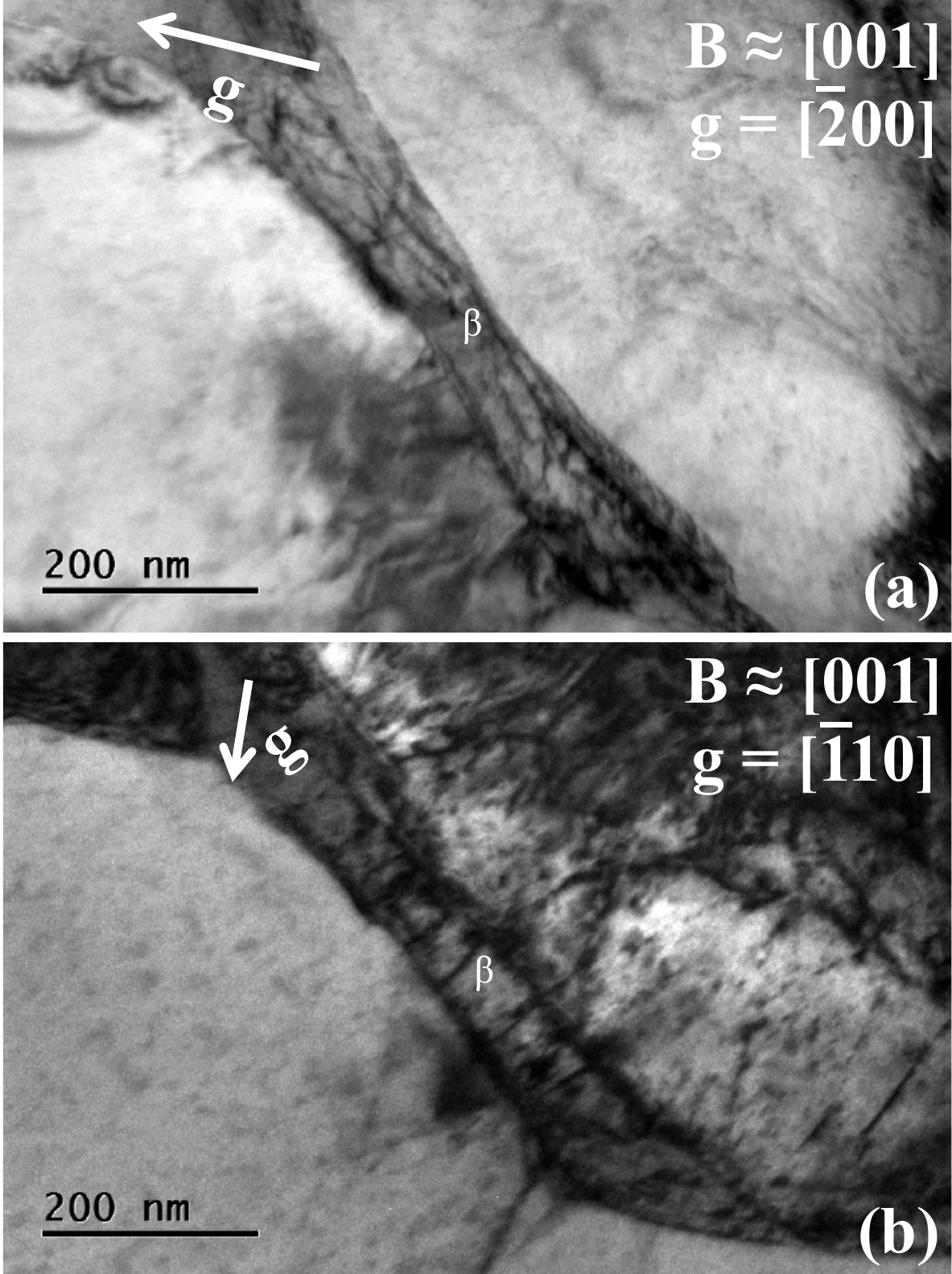}\end{center}
\caption{BF-TEM showing dislocations in ligament $\beta_2$ of a two phase region $\alpha_s+\beta(2)$ in foil 2 under two-beam conditions near $B \approx [001]$: (a) network of $b_{\bar{1}11}$  and $b_{11\bar{1}}$ dislocations under $g =[\bar{2}00]$ and (b) $b_{010}$ dislocations under $g =[\bar{1}10]$.}
\end{figure}

	 \section{Discussion}
\textcolor{black}{We will first discuss the dislocation behaviours found in both near-origin faceted fatigue crack growth and in the initiating grain, and compare to the observations made previously for similarly-oriented grains from the gauge section~\cite{joseph2018slip}. Then, we turn to examine slip transfer in the $\alpha_s+\beta$ regions, before finally turning to the initiation grain to ask what is different about the dislocation configurations associated with crack nucleation.} 
     
	 \subsection{Dislocation interactions in $\alpha_p$ grains}
	 Isolated dislocations are rarely observed in this alloy, indicating that dislocation groups or arrays are responsible for deformation. The two foils investigated exhibit strongly planar slip in the $\alpha_p$ grains; due to the lack of forest hardening, this kind of planar slip leads to a low rate of work hardening. Planar slip in Al-containing hcp $\alpha$-Ti is believed to be a consequence of the short range ordering of Ti and Al atoms \cite{savage2002observations}. The passage of a single dislocation through $\alpha_2$-Ti$_3$Al precipitates, possibly in pairs, makes it easier for subsequent dislocations to traverse the same plane, leading to pile-up in planar slip bands.  The dense pile-ups observed indicate that, once the ordered precipitates / short range ordered regions are sheared, they no longer present a barrier to dislocation motion. Hence the local hardening rate for a slip plane is small once activity commences. 
	 
	 The pile-ups in $\alpha_p(1)$ of foil 1 near crack nucleation are dense, thick, double ended and non-classical, \textcolor{black}{Figure 3 a and b}. Following Tanaka and Mura~\cite{tanaka1981dislocation}, the intersection of this kind of large thickness pile-up with a microstructural boundary can trigger microcrack initiation. This is evidenced by the shearing of $\beta$ plates by the intense slip bands, Figure 3e. As these are not classical pile-ups, extensive run-back of dislocations contained within the arrays does not occur. This will then lead to high stress concentration on boundaries, due to the stress generated by these dislocations. The pile-ups in foil 2 are also dense and found to be double-ended, \textcolor{black}{Figure 7}. The motion of dislocations in opposite directions by multiple cross-slip events has been observed in $\alpha$-Ti during in-situ tensile testing \cite{kacher2016situ}, which provides a mechanism of formation of a double ended pile-up. We have previously observed this kind of dislocation multiplication by multiple cross-slip events, leading to double ended pile-ups, in TEM foils from the gauge section of this sample \cite{joseph2018slip}. 
	 
	 The various slip systems in grain $\alpha_p$(1) in foil 1 were not observed to interact strongly, except the jogging of isolated straight dislocation lines, \textcolor{black}{Figure 3c}. These are jogs in edge dislocations, which can move with the parent dislocation as the Burgers' vector of the jogs lie in the slip plane of the parent dislocation. However, a higher Peierls stress is still required to move such jogged dislocations \cite{gumbsch2011multiscale}. 
     
	 In foil 1, the $\alpha_p$ grain possessing a double ended pile-up had no neighbouring $\alpha_p$ grains, instead being bounded (in the section observed) by two-phase regions. Because of its higher critical resolved shear stress (CRSS), it can be difficult to activate $\langle c+a \rangle$ pyramidal slip. It's observation in the present case can be rationalised as being a consequence of constraint by the adjoining two phase regions. The pile-ups in this grain did not transfer across the boundary and it was also observed that the mixed components of the straight dislocations also piled-up, Figure 3d. The nature of these long dislocations with edge and mixed components suggests that these dislocation loops originated from the $\alpha_p /\alpha_s + \beta$ boundary. Thus, these boundaries act both as dislocation sources as well as providing resistance to slip transfer.   
	 
	 Many instances of slip transfer were observed between the similarly oriented $\alpha_p$  grains in foil 2, \textcolor{black}{Figure 6}. \textcolor{black}{It can be seen from this figure that} both the dislocations in the pile-ups and single dislocations were found to transfer across the boundaries; pile-up transferring across the boundary in grain pairs 1/2 and 2/3, and single dislocation transfer between grains 3 and 4. This suggests that the large stress concentrations associated with pile-ups are not required for slip transfer across such similarly-oriented grains. Furthermore, such grain boundaries offer little resistance to dislocation motion. Among the many pile-ups, some traversed an entire grain, indicating that those are the primary slip activated. Subsequently, once stress is increased by a primary pile-up, secondary slip systems can be activated, \textcolor{black}{Figure 9}. 
	 
	 The observation of dislocation loops originating from the $\alpha_p /\alpha_s + \beta$ boundary in Figure 10b also supports that this boundary acted as a dislocation source. Among the five $\alpha_p$  grains investigated in foil 2, only one, $\alpha_p$(5), was found to deform homogeneously; this grain's $c$-axis was oriented $32\degree$ to the loading direction whereas all the other grains were inclined by $>45\degree$, \textcolor{black}{Table 2}. $\alpha_p(2)$ in foil 1 also showed homogeneous deformation, \textcolor{black}{Figure 5}, with its $c$-axis $4\degree$ from the loading direction. This suggests that it is relatively rare to observe pile-ups in `hard-oriented' grains, those with their $c$-axis near the loading direction.
	 	 
	 \subsection{Dislocation interactions between $\alpha_s$  and $\beta$ plates}
	 From the dislocation structures observed in the two-phase regions, sec 3.2.2, the following mechanism is suggested for the deformation of the $\alpha_s /\beta$ structure. The $\alpha_s$  and $\beta$ plates follow the Burgers orientation relation, $(101)_\beta || (0001)_\alpha$ and $[1\bar{1}\bar{1}]_\beta || [2\bar{1}\bar{1}0]_\alpha$, Fig. 13. The crystallographic relations bring one of the hcp $\langle 2\bar{1}\bar{1}0 \rangle$ close packed directions into near-coincidence with the bcc $\langle 1\bar{1}\bar{1} \rangle$ slip direction. A deviation from the Burgers' orientation relation exists between individual $\alpha$ and $\beta$ plates, with a $0.7\degree$ between $a_1$ and $b_1$, and a misorientation of $11.1^\circ$ observed between $a_2$ and $b_2$ \cite{suri1999room,savage2001deformation, savage2002observations}. The third $a$-type slip direction $a_3$ in the $\alpha$ plate does not have a closely aligned $\langle 111\rangle_\beta$ direction but is relatively closely aligned with $a[010]$. The misorientation between $a_3$ in the $\alpha$ plate and $a[010]$ in $\beta$ ligament is $5.96\degree$ and the mismatch in their lengths is $8.9\%$ \cite{savage2004anisotropy}.
	 
	The schematic in Fig. 16 shows a possible mechanism of deformation in the two-phase region. Initially the $a_3$ dislocation loops nucleate from the $\alpha_s/\beta$ boundary, gliding on basal planes. These loops then slip across the $\beta$ ligament to the neighbouring $\alpha_s$ plate. \textcolor{black}{This is expected since the slip easily traverses across the $\beta$ ligaments when they are very thin. Zhang et al \cite{zhang2016determination} suggested that this slip transmission across the boundary is favoured by closely aligned slip plane normals in the two phases and the global Schmid factors are not appropriate to identify the local slip initiation in either phase.}  
    
    The one-to-one correspondence between $\langle a_3 \rangle$  dislocations in all three $\alpha_s$ plates investigated shows that these dislocations transferred across the $\beta$. These are the edge components of the dislocation loops. During this slip transfer process, the $\beta$ plates are also sheared, resulting in $a[010]$-type dislocations. Since $a_3$ direction in $\alpha$ plate has no closely matched $(a/2)\langle 111 \rangle$ slip direction in the $\beta$ plate, this relatively closely aligned $a[010]$ slip is expected.  The following reaction could occur during slip transmission across the $\alpha_s/\beta$ boundary
	\begin{equation} a_3 \rightarrow b_{010} + r_{a_3}\end{equation}
\begin{figure}[t]
\begin{center}\includegraphics[width=45mm]{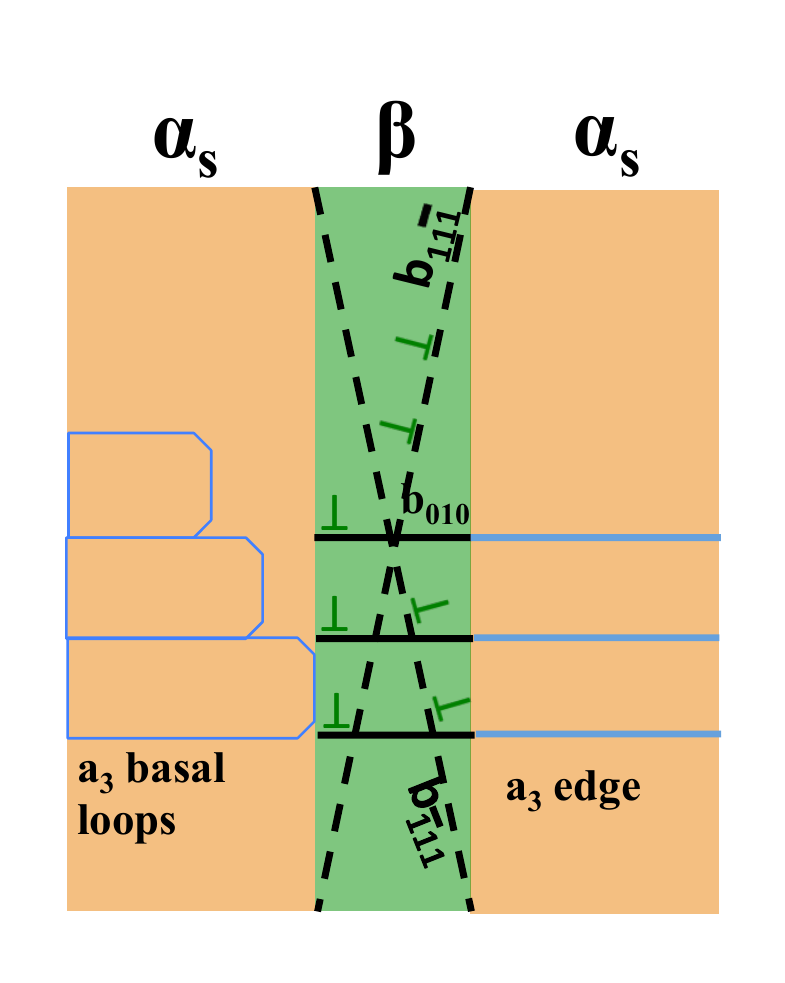}\end{center}
\caption{Schematic showing an interpretation of the dislocation interactions observed in the $\alpha_s$ and $\beta$ plates in two-phase region $\alpha_s+\beta(2)$ of foil 2.}
\end{figure}
    where $r_{a_3}$ is a residual dislocation resulting from the misorientation and mismatch between the $a_3$ and $ b_{010}$ dislocations. Such residual matrix dislocations can be observed near the $\alpha_s/\beta$ boundary, Figure 14d.  The transmission of $a_3$ dislocations into a $\beta$ plate requires higher stresses since the $a_3$ direction in the $\alpha$ doesn't have a closely aligned $\langle 111 \rangle$ direction in the $\beta$. However, the high interfacial stress at the $\alpha_s/\beta$ interface might have caused this transmission. \textcolor{black}{Further, the slip strength of $\beta$ plates was found to be of the same order as that of basal slip in $\alpha$ plates which suggest that the $\beta$ plates may not act as a significant barrier for slip transmission \cite{zhang2016determination}.}
    
    Glide of such $a\langle 010 \rangle$ dislocations is rarely reported since the mobility of $a \langle 010 \rangle$ dislocations are much lower than that of the favoured $(a/2)\langle 111 \rangle$ dislocations. $\langle 010 \rangle$ type slip activity has been reported in some bcc metals, \emph{e.g.} in lithium \cite{gorgas1986slip} and niobium \cite{reid1966twinning}. Furthermore, Savage et al \cite{savage2004anisotropy} observed this $a\langle 010\rangle$ dislocations in colony-structured Ti-6242 alloy after room temperature compression. Such a $b_{010}$ dislocation in the $\beta$ will then transfer into an $a_3$ dislocation in the next  $\alpha$ plate as 	
    \begin{equation}  b_{010} \rightarrow a_3 + (-r_{a_3})\end{equation}	
	The two possibilities for the existence of these $a\langle 010\rangle$ dislocations are either (i) that the intersection of $b_1$ and $b_2$ dislocations might have resulted in the formation of this dislocation, as described in Nb~\cite{louchet1975dislocation}, or (ii) they could result from the direct slip transmission of $a_3$ dislocations in $\alpha$ to $b_{010}$ dislocations in $\beta$. The one-to-one correspondence between the $a_3$ dislocations in $\alpha_{s}$ and $b_{010}$ dislocations in the $\beta$ supports the second mechanism. 
	
	The network of $b_1$ and $b_2$ dislocations observed in the $\beta$ plate we then expect to result from the dissociation of these $a\langle 010 \rangle$ dislocations into $b_1$ and $b_2$ dislocations
    \begin{equation} a[010] \rightarrow (a/2)[11\bar{1}] + (a/2)[\bar{1}11]\end{equation}	
    This decomposition is favored under high stresses due to the much higher mobility of $b_1$ and $b_2$ dislocations than $b_{010}$ dislocations\cite{savage2004anisotropy}; the high stresses will arise from  the applied shear stress and the interfacial stresses naturally present between the phases as a result of precipitation and cool-down from the formation temperature.
    
    \textcolor{black}{Thus the $\beta$ plates do not appear to form an impenetrable barrier to slip as observed in \cite{zhang2016determination}. An another important observation was made in the work of Zhang et al \cite{zhang2016determination} where the $\beta$ plates were found to show the strongest rate sensitivity, which is important in rate dependent deformation such as fatigue with a load hold.}

	\subsection{Crack nucleation mechanism}
	Generally, it is believed that fatigue crack initiation in Ti alloys occurs by facet formation~\cite{sinha2006observations, bache1997electron,evans2003dwell,  kirane2008cold, dunne2008mechanisms, anahid2011dwell, zheng2016dwell} and that the facets correspond to $\alpha_p$ grains. Our fractographic analysis also showed facet formation near the crack  initiation site with the appearance of $\alpha_p$ grains. However, after FIB foil extraction, some of the facets are found to be two phase regions. This is possible when slip in $\alpha_{s}$ plates transfers across the $\beta$ along the closely aligned directions, \emph{i.e.} $\langle a_1 \rangle$ and $\langle a_2 \rangle$ directions in $\alpha_{s}$ into $\langle b_1 \rangle$ and $\langle b_2 \rangle$ in the retained $\beta$. In that case, the two phase region can be featureless, with the appearance of a smooth facet. This kind of facet with an underlying two phase microstructure has on occasion been observed, \emph{e.g} by Pilchak et al. \cite{pilchak2010crystallography} during the fatigue of fully lamellar Ti-6Al-4V.    
	
	Based on the observations in foil 1 near the fatigue crack initiation, the following crack nucleation mechanism is proposed. Near the crack initiation, substantial dislocation activity was observed in grain $\alpha_p(1)$, which is adjacent to a transformed $\beta$ region on either side, \textcolor{black}{Figure 3a}. The sketch in Figure 17a shows the overall dislocation activity in this grain.  The dislocations were arranged as a double-ended pile-up containing dislocations of opposite sign in the successive (stacked) pyramidal planes.  This suggests that during the first half of the loading cycle, pile-up of dislocations with a positive sign occurs on the first slip plane. Then, reverse plasticity in the second half cycle causes pile-up of dislocations of negative sign on successive planes close to the first slip plane. It is also possible for the dislocations in the first plane to reverse completely during the second half of the cycle. If such complete reversal of piled-up dislocations in the first plane occurs, then there is no accumulation of dislocations as a double ended pile-up \textcolor{black}{\cite{tanaka1981dislocation}}. Incomplete reversibility of dislocation motion could be realized if the lattice friction stress against dislocation motion \textcolor{black}{is higher in the reverse direction than in the forward direction \cite{tanaka1981dislocation}. This could be rationalised as follows.}  
\begin{figure*}[t]
\begin{center}\includegraphics[width=145mm]{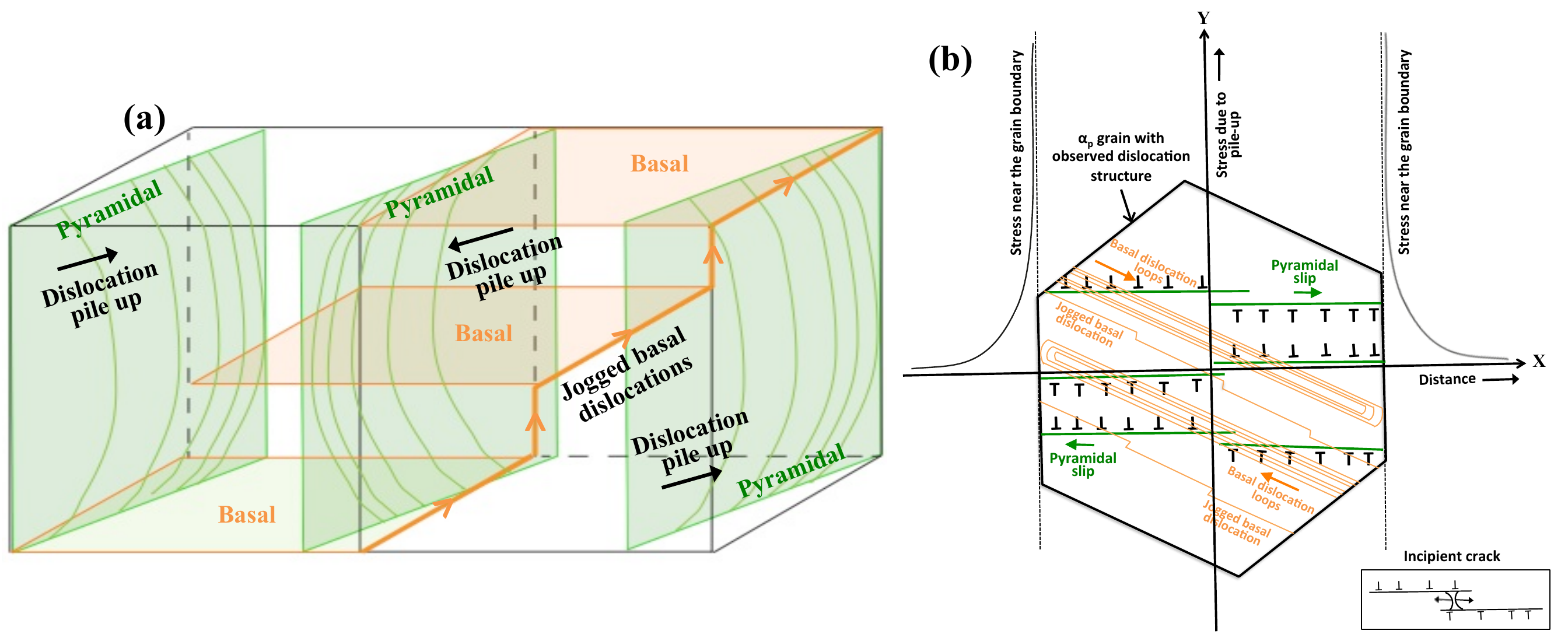}\end{center}
\caption{Schematic showing (a) the dislocation arrangements in a fracture initiating $\alpha_p$ grain as a double ended pile-up and (b) the stress distribution \textcolor{black}{near the grain with this double ended pile-up}. Incipient crack between the slip planes having opposite dislocations along the dislocation lines \textcolor{black}{is shown in the inset of (b).}}
\end{figure*}
    
The back stress developed by these dislocations is negative in the vicinity of pileup in the first plane, which would cause plastic flow in the reverse direction in the neighbouring plane during subsequent reverse loading \textcolor{black}{\cite{tanaka1981dislocation}}. The pile-up of negative dislocations on the second plane then causes a positive back stress on the first plane. This back stress enhances the pile-up of positive dislocations during the next stage of forward loading.  As the number of cycles increases the number of dislocations in each slip plane as well as the number of piled-up planes increases, hence resulting in the formation of a double ended pile-up, similar to the dislocation distribution near a crack subjected to a shear stress \cite{bilby99dislocations, tanaka1981dislocation, lothe1982theory}. This kind of pile-up is possible only when the dislocation sources are located very close to each other, \textcolor{black}{because for an isolated dislocation source, the dislocation motion on a plane becomes reversible as there is no  back stress produced by the dislocations on a neighbouring plane.} Therefore, the reversibility or otherwise of dislocation motion will depend on the statistical distribution of dislocation sources.
	
	An alternative explanation for the formation of the double ended pile-up might be the occurrence of multiple cross-slip events, as observed in the grains from the gauge section of the same sample \cite{joseph2018slip}. But, the dislocation debris and loops that would be produced by such cross-slip events were not observed in this grain, suggesting that instead that the irreversibility mechanism outlined above was responsible for the double ended pile-up in this grain. 
	
	The crack might then nucleate from the accumulated dislocation pile-ups in the following ways: (i) along the slip band \cite{lin1986fatigue, ahmed2001electron}; (ii) from the grain boundary or two phase region boundary where the slip bands intersect \textcolor{red}{\cite{tanaka1981dislocation}}; or (iii) the stress concentration due to this pile-up can nucleate a crack in the neighbouring grain due to load shedding~\cite{stroh1954formation, evans1994dwell,britton2012stress}. The first two cases were not observed in the present study since the crack was not observed along the direction of slip bands or along the grain or two-phase region boundary. Case  (iii) is also not possible in this case as this grain with dislocation pile-ups did not possess adjacent alpha grains and is bounded by two phase regions. Further, these two phase $\alpha_s+\beta$ regions are found to offer high resistance for the slip transfer of piled-up dislocations from $\alpha_p$ grains \cite{joseph2018slip}. 
	
	Instead, we suggest that the embryonic crack might have nucleated from this double ended pile-up as follows.  The stress field in this case would be the sum of applied stress and the internal stresses due to the dislocation pile-ups on successive slip planes. In addition, the mixed components of the basal dislocation loops were also found to pile up near the boundary \textcolor{black}{(Figure 3b)}, causing further stress intensification. 
    
    \textcolor{black}{The microplasticity produced by the slip bands within individual grains has been considered as an important parameter in microstructure sensitive computational methods by D.L. McDowell et al. \cite{mcdowell2010microstructure, przybyla2011simulated, przybyla2013microstructure}. It is found to be a key driving force for fatigue crack nucleation and microstructurally small fatigue crack growth. The accumulated plastic strain can be used as a fatigue indicator parameter (FIP) for crack formation. This parameter was found to be an good indicator for predicting fatigue crack nucleation in both low cycle and high cycle fatigue.} 
    
    The stress distribution due to this double ended pile-up \textcolor{black}{together with the dislocation structures in the $\alpha_p$ grain} is shown in the schematic, Figure 17b. The stress intensification takes place near the tips of the pile-up. A large tensile stress is built up between the two planes of opposite sign dislocations in the pile-up, nucleating the facet. The tensile stress at half the distance between the two slip planes is given as \cite{tanaka1981dislocation}:
\begin{equation}
\sigma_{xx} =3\sqrt{a} \usk n(\Delta\tau-2\kappa)/\sqrt{2h}
\end{equation}
	where 2a is the grain size, n is the number of stress cycles, $\Delta\tau$ is the cyclic shear stress applied to the grain, $\kappa$ is the lattice friction stress and $h$ is the slip plane interplanar spacing. 
    
    The tensile stress would split the atomic planes along the dislocation lines \cite{fujita1958dislocation}, inset of Figure 17\textcolor{red}{b}. This can be observed in Figure 3a -- the crack grew along the dislocation line direction. Therefore the large tensile stress resulting from the double ended pile-up appears to be the mechanism responsible for crack nucleation in the $\alpha_p$ grain.  From the TEM foil, it was observed that the fracture plane is found to be oriented 5--$8\degree$ from the basal plane. This near-basal plane cracking rather than true basal plane cracking might be a consequence of the observed jogging of the basal dislocations. 
    
    \textcolor{black}{A new microstructural level critical stored energy density criterion proposed by Wan et al. \cite{wan2014stored}  argued that the statistically stored and geometrically necessary dislocations played a role in developing highly localized stress states, capable of causing localized cleavage. The critical plane criterion with strain based damage parameter~\cite{kandil1982} showed that the damage was controlled by the shear strain on the crystallographic slip plane and the fatigue crack growth was assisted by the normal strains on the same planes.}
    
    \textcolor{black}{The fatigue indicator parameter (FIP) suggested by C.P. Przybyla et al. \cite{przybyla2013microstructure} is formulated to capture the effect of crystallographic orientation of primary $\alpha$ grains on fatigue crack formation in $\alpha+\beta$ Ti alloys. By design of the parameter, grains oriented favourably for slip on $\langle c+a\rangle$ pyramidal show an extreme value of this FIP,  similar to the present observation of crack nucleation from a primary $\alpha$ grain with a $\langle c+a\rangle$ double ended pile-up in pyramidal planes.}

	\section{Conclusions}
    The dislocation structures associated with the fatigue crack initiation site in aged bimodal Ti6242Si subjected to low cycle fatigue were examined, and compared to those previously found in the gauge and to those found in grains associated with crack growth. The following conclusions are drawn.
    
	1. The alloy mainly deforms by cooperative movement of a large number of dislocations in the primary alpha grains, which predominantly deform by planar slip; interaction between the different operating slip systems is relatively rare.  These pile-ups are found to be localized within the grain when the grain shares its boundary with two-phase regions or dissimilarly oriented grains; the slip bands transfer across the boundary when the grain has similarly oriented neighbour. The boundary between the primary alpha grain and a two phase region acts both as a dislocation source and as a barrier to \textcolor{black}{slip transfer between primary alpha grains}.
	
    2. Dislocations in two phase $\alpha+\beta$ regions were observed to nucleate from the $\alpha$/$\beta$ boundaries as dislocation loops. A number of basal $\langle a \rangle$ loops were observed on basal planes in the $\alpha_s$ plates; the edge component of these loops were found to transfer across the retained $\beta$ ligaments. As the $a_3$ slip direction in the $\alpha_s$ plates doesn't have a closely aligned slip direction in the $\beta$, direct transmission needs higher stress. These $a_3$ dislocations become $a[010]$ dislocations in the $\beta$ and subsequently decompose into a pair of mobile $(a/2)\langle 111\rangle$ dislocations, observed as dislocation networks in the $\beta$.
	
	3. The high dislocation density observed in one of the primary alpha grains nearest the initiation site suggests that crack nucleation occurred from this grain. $\langle c+a\rangle$ pyramidal slip was observed in this grain, which may be a consequence of the constraint imposed by the $\alpha_s+\beta$ two phase regions on the either side of this grain, as well as the applied stress. Dislocations were observed to be piled up on both the ends of the grain, with positive and negative dislocations observed on neighbouring slip planes. Such structures can result from \textcolor{black}{the incomplete reversibility of dislocation motion due to a slightly higher friction stress in the reverse direction than in the forward direction}. This kind of double ended pile-up results in a large tensile stress between the two planes, providing a mechanism of crack initiation by plane splitting along the dislocation lines. The observation of superjogs on the basal dislocations provides a rationale for why such cracks nucleate near rather than on the basal plane.    

\section*{Acknowledgements}
\noindent\small The authors wish to acknowledge the contribution and useful discussions with Prof D Rugg at Rolls-Royce plc, Prof FPE Dunne and Dr TB Britton at Imperial, and Prof AJ Wilkinson in Oxford, with whom we are funded under the Hexmat EPSRC programme grant EP/K034332/1.

\section*{References}

	\bibliographystyle{unsrt}
\small\setlength{\itemsep}{0cm}	\bibliography{sliptransfer2_090418}
	
\end{document}